\documentclass[10pt, amsmath, amssymb, aps, prd, showpacs, showkeys]{revtex4-2}

\usepackage{xcolor}

\usepackage{graphicx}
\usepackage{dcolumn}
\usepackage{bm}
\usepackage{hyperref}

\begin{document}

\preprint{APS/123-QED}

\title{\textbf{Gravitational foundations and exact solutions in $n$--dimensional fractional cosmology}}
\author{S. M. M. Rasouli}
 \altaffiliation{Departamento de F\'{i}sica,
Centro de Matem\'{a}tica e Aplica\c{c}\~{o}es (CMA-UBI),
Universidade da Beira Interior,
Rua Marqu\^{e}s d'Avila
e Bolama, 6200-001 Covilh\~{a}, Portugal}
 \email{mrasouli@ubi.pt}

\author{J. Marto}
 \altaffiliation{Departamento de F\'{i}sica,
Centro de Matem\'{a}tica e Aplica\c{c}\~{o}es (CMA-UBI),
Universidade da Beira Interior,
Rua Marqu\^{e}s d'Avila
e Bolama, 6200-001 Covilh\~{a}, Portugal}
\author{D. A. V. Oliveira}
 \altaffiliation{Departamento de F\'{i}sica,
Centro de Matem\'{a}tica e Aplica\c{c}\~{o}es (CMA-UBI),
Universidade da Beira Interior,
Rua Marqu\^{e}s d'Avila
e Bolama, 6200-001 Covilh\~{a}, Portugal}

\author{P. Moniz}
 \altaffiliation{Departamento de F\'{i}sica,
Centro de Matem\'{a}tica e Aplica\c{c}\~{o}es (CMA-UBI),
Universidade da Beira Interior,
Rua Marqu\^{e}s d'Avila
e Bolama, 6200-001 Covilh\~{a}, Portugal}

\date{13June2026}
\begin{abstract}
Three theoretically plausible techniques to developing a fractional scalar field cosmological model are pointed in this paper; the time-dependent kernel weighted action being then selected. Upon this choice, we proceed to establish (i) a time weighted action associated with the generalized scalar field cosmology; and (ii) a fractional cosmological model in $n$ dimensions considering the FLRW metric and a generalized version of the S\'{a}ez-Ballester (SB) theory. Our study focuses on the following purposes. Firstly, to investigate the fundamental gravitational structural features of the model, we analyze the dynamical behavior of the field equations, the fulfillment of the Bianchi identities, the associated conservation laws, and the application of the second Noether theorem at the background and first-order perturbation levels. Moreover, the model's distinguishing characteristics and theoretical differences from the corresponding standard scenarios are also investigated.
	Secondly, we aim to obtain exact analytical solutions and analyze the time evolution of key cosmological quantities, considering the fractional parameter effects. Furthermore, the model's predictions are compared with those of the corresponding standard models.
	Lastly, we propose new ideas to further generalize our model, with a focus on constructing an effective potential and investigating the conditions under which bounce solutions may emerge.
   \end{abstract}


\maketitle
\section{Introduction}
\label{SecI}

Gravitational theories in higher dimensions, based on their essential motivations, have evolved into various cosmological models that aim to describe the evolution of the universe. In particular, numerous frameworks have been developed to address outstanding problems related to gravity and cosmology \cite{Salgado:2002qy,Liko:2003ha,Rasouli:2014sda,Rasouli:2016ngl,troisi2017higher}. Notable among these higher dimensional frameworks are conventional Kaluza--Klein (KK) theory \cite{appelquist1983quantum,Overduin:1997sri}, supergravity, string theory, and M-theory \cite{moniz1998origin,moniz2010semiclassical,Bojowald:2010cj,Taylor:2011wt,Sen:2024nfd,Blair:2023noj}, and contemporary advancements of KK frameworks \cite{doroud2009class,Rasouli:2009rs,Maartens:2010ar,Atazadeh:2014joa,Rasouli:2022tmc}.

 Scalar fields have been essential in the development of modern physics. In gravitational theories, particularly when introduced into the Einstein--Hilbert (EH) action, where they can couple to gravity either minimally \cite{Cruz:2000ds,paliathanasis2015dynamical,stavrinos2025modified} or non--minimally \cite{Clifton:2011jh,gashti2025noncommutativity,ildes2023analytic,sadeghi2024swampland,velasquez2024jordan}, broadening the scope of general relativity and facilitating significant progress in our understanding of the universe, particularly in the semi-classical regime \cite{Guzman:2007zza,Burgazli:2015mzm,rasouli2014gravitational,leon2022scalar,Lopez-Picon:2023rhc}. One of the scalar field cosmological models that has recently been employed to describe the evolution of the universe is the S\'{a}ez-Ballester (SB) theory \cite{saez1986simple,Socorro:2009pt,jamil2012bianchi,Do:2025xhm}, which will be the focus of the present work.
 
In the SB framework, a scalar field with a non--canonical kinetic term involving two constant parameters is minimally coupled to the geometry \cite{saez1986simple, mishra2020cosmological, vinutha2022study,chetia2023particle,luciano2023saez}. However, by applying appropriate transformations (even in the presence of a scalar potential), it can be easily shown that this model can be related to the standard minimally coupled scalar field (MCSF) model \cite{Rasouli:2022tjn,Rasouli:2023mae}, as will be discussed in the next section. 
In recent years, the SB model has been applied in various approaches to address a range of issues in gravitational and cosmological contexts \cite{singh2022vacuum,rasouli2023noncompactified,singh2024observational,chokyi2024cosmology,do2025anisotropic}. However, the main reason for selecting a generalized version of this model is its sophisticated yet broadened framework compared to the other MCSF models \cite{jamil2012bianchi,mishra2021bianchi,garg2021cosmological,daimary2022five}. Concretely, the two adjustable parameters present in the non--canonical kinetic term, along with the fractional parameter introduced in our model, offer the necessary flexibility to obtain the generalized solutions.

The fundamental mathematical framework for most fields of science and engineering has historically been based on classical differential and integral calculus. Nevertheless, in the past few decades, fractional calculus (which extends classical calculus by allowing derivatives and integrals to have non-integer or even complex orders) has emerged as a valuable tool for investigating previously unresolved issues and modeling intricate phenomena. Physics is similarly influenced by this development; see, for 
instance, \cite{riewe1996nonconservative,hilfer2000applications,calcagni2012geometry,uchaikin2013fractional,el2013non,laskin2019nonlocal,calcagni2021multifractional}, references therein and related investigations. (While fractional derivatives with complex orders are mathematically permissible, practical applications in physics typically focus on real orders.) Fractional calculus has shown significant utility not only in classical regimes \cite{costa2023estimated,Rasouli:2024crg,el2024schwarzschild,Jalalzadeh:2024qej,Socorro:2024poa,micolta2025fractional} but also within various quantum frameworks \cite{calcagni2010quantum,calcagni2010fractal,Rasouli:2021lgy,socorro2023quantum,Varao:2024eig,Duan:2025ovv,Umar:2025ppo}. Among the most widely used tools in these contexts is the implementation of fractional derivatives, such as the Riemann--Liouville and Caputo fractional derivatives, whose use depends on the specific requirements of a given scenario under consideration \cite{odibat2006approximations,roberts2009fractional,li2011riemann,torres2020quantum,calcagni2021classical,shchigolev2021fractional,marroquin2024conformal}.

A pertinent question in this context is how a given gravitational theory can be consistently reformulated within a fractional framework. From a general perspective, different routes have been explored in the literature  see for instance 
\cite{Roberts:2009ix,Shchigolev:2010vh,Jamil:2011uj,Micolta-Riascos:2023mqo,calcagni2021multifractional,calcagni2023ultraviolet,briscese2026fractal} and the references therein. One of the most direct approaches consists in modifying the integration measure by introducing non-trivial weights or kernels, leading to weighted actions that can effectively capture scale-dependent behavior and memory effects in the dynamics 
\cite{shchigolev2013fractional,calcagni2010quantum,garcia2022cosmology,Rasouli:2026uax}.
Alternatively, fractional or non-local modifications may be implemented at the level of certain differential operators, either in the action or directly in the field equations, while keeping the standard integration measure unchanged \cite{shchigolev2021fractional,Rasouli:2021lgy,calcagni2021classical,Micolta-Riascos:2023mqo,Rasouli:2024crg,benetti2025relativistic}.
In some frameworks, it is conceivable that both of types of modification can be combined, so that the integration measure and the differential operators are simultaneously generalized, potentially leading to a richer description of non-local and scale-dependent phenomena.
At a more speculative level, one may also envisage scenarios in which the underlying geometric structure of spacetime is effectively modified, for instance, through conceptual extensions of geometric quantities such as the connection or the Riemann tensor. Although such directions are still under development, they may provide complementary avenues for incorporating fractional features into gravitational theories.

The present work is primarily situated within the first class, since the fractional modification is introduced through a weighted action. However, it is important to emphasize that these categories are not mutually exclusive, and several models in the literature combine elements from more than one class. Our use of this classification is therefore intended only as a heuristic guide to organize the discussion, rather than as a claim of a new or exhaustive classification.

Each of the mentioned strategies introduces distinct features and advantages compared to their classical counterparts, while also presenting specific challenges and complexities in interpretation.

In this work, let us implement the first and simplest strategy. We establish a fractional SB cosmological model in $n$ dimensions by incorporating a Riemann--Liouville type kernel into the time integral of the action. Our objectives are not only to investigate the gravitational structure of the proposed framework but also to obtain exact solutions and analyze their cosmological predictions in comparison to those of the corresponding standard models.

In addition, in the original SB framework, no scalar potential was considered \cite{saez1986simple}. However, in this work, we extend the theory to higher dimensions by incorporating a scalar potential alongside a time-dependent kernel in the action, effectively yielding a generalized fractional scalar field cosmology. This constitutes the main significant differences. 
Our main objectives are then to address the following questions.
\begin{itemize}
\item

\textbf{Structural distinctions and consistency conditions:}
What structural aspects distinguish this model from its standard (non-fractional) counterparts?  What conditions allow this model to satisfy time diffeomorphism invariance, the Bianchi identities, and the second Noether theorem \cite{baez1994gauge,kosmann2010noether} not only at the background level but also at first order perturbations?  How do fractional field equations differ fundamentally from their conventional analogs?

	\item
 
\textbf{Cosmological dynamics and analytical tractability:}
Can this fractional cosmological model describe all epochs of the universe? Is it possible to obtain exact solutions without assuming specific potentials? Alternatively, can one, like standard counterparts, still apply ad hoc potentials for solving the field equations? How can we develop dynamical systems for such a fractional model? Are generalized dynamical systems similar to those used in standard situations appropriate for studying a completely generic potential in this fractional framework?
\end{itemize}

In the present work, our main motivation is to investigate certain gravitational foundations of fractional cosmological models, including the structure of the field equations, the role of the Bianchi identities, and symmetries related to Noether's theorem. In order to investigate these aspects within a sufficiently general framework, we consider a relatively general action that can accommodate several classes of modified cosmological scenarios.

In addition, the framework adopted here naturally extends a line of research in which different aspects of the SB theory have been explored in various contexts, including higher--dimensional realizations of the model. In particular, previous studies have analyzed configurations such as the SB theory in $n$--dimensional spacetimes \cite{Rasouli:2017glb,Rasouli:2019axn,Rasouli:2022hnp}. The present work aims to further develop this direction by incorporating fractional weighting of the action and investigating its consequences for the background cosmological dynamics and the corresponding exact solutions. In this way, the model considered here provides a unified setting to study the interplay between scalar-field dynamics, fractional effects, and spacetime dimensionality.

Furthermore, the proposed framework admits several limiting cases that reduce to previously studied models. For instance, in the limit $\alpha = 1$, the fractional structure disappears and the model reduces to the standard $n$--dimensional SB theory. In addition, by switching off one, two, or even all of these modifications, one can recover more specific scenarios, which serve as useful consistency checks for the proposed model.

This study is organized as follows.
In the next section, we present an extended version of the SB theory with a scalar potential in $n$-dimensional spacetime and review one of its important features.The model is further generalized by incorporating a time-dependent kernel.
In Section \ref{Pert-Dyn}, we derive the governing equations for first-order perturbations associated with the time-weighted scalar field cosmology established in Section \ref{SecII}. In particular, we obtain the cosmological equations corresponding to the components of the Einstein tensor, together with the perturbed Klein-Gordon equation in the presence of the time-dependent kernel and the generalized Mukhanov–Sasaki equation, and discuss their main differences from the standard case. 
In Section \ref{FLRW-SB-Cosmology}, considering the FLRW metric, we establish a fractional SB cosmological framework and analyze its main features, and contrast its strengths and shortcomings with those of its standard (non-fractional) counterparts. Moreover, we obtain exact analytical solutions without imposing ad hoc assumptions, which are applicable to arbitrary scalar potentials. 
In Section \ref{behav-alpha}, we investigate the time evolution of the cosmological quantities through numerical analysis in two distinct regimes.
In Section \ref{SecIV}, we formulate two different representations of dynamical systems of the fractional SB model.
Finally, in Section \ref{SecV}, we summarize our results and outline possible directions for future research.


\section{Generalized scalar field cosmology in arbitrary dimensions: a time-weighted action formulation}
\label{SecII}

In this section, we establish a generalized framework for scalar-field cosmology (with a non--canonical kinetic term), in which a time-dependent kernel is systematically incorporated into the structure of the action. Throughout this paper, by \textit{scalar field cosmology} we refer to models in which the scalar field is minimally coupled to gravity.

Since the  S\'{a}ez--Ballester (SB) cosmological model represents a particular realization of the mentioned framework and will be the main focus of the sections \ref{FLRW-SB-Cosmology}--\ref{SecIV}, it is appropriate to first introduce its formulation in $n$ dimensions.

In the standard SB framework, the scalar field that features a generalized non-canonical kinetic term, is minimally coupled to gravity. In this model, the ordinary matter sector was included, but no potential term was considered \cite{saez1986simple}. Cosmological models constructed within this setup can exhibit rich dynamics in the presence of ordinary matter or various potential forms. However, such models often rely on phenomenological assumptions regarding potential and matter content to describe the evolution of the universe \cite{Singh:2023evc,Do:2025xhm}.

Let us first briefly introduce a generalized version of the SB model and highlight one of its key properties, namely its equivalence, under a suitable field redefinition, to the Einstein--Hilbert action with a minimally coupled scalar field. Such an action in $n$ dimensions can be written as
\begin{eqnarray}\label{SB-action-standard}
S_{_{\rm SB}}=\int d^{n}\!x {\cal L}_{_{\rm SB}}=\int  d^{n}\!x \sqrt{-g}\,
 \Big[\frac{1}{2\kappa_n}R^{^{(n)}}-\frac{1}{2}{\cal J}(\phi)\, g^{\alpha\beta}\,({\nabla}_\alpha\phi)({\nabla}_\beta\phi)
 -V(\phi)\Big],
\end{eqnarray}
where $\kappa_n \equiv8\pi G_n$ and ${\cal L}_{_{\rm SB}}$ denote the Lagrangian density of a modified SB gravitational model in arbitrary dimensions \cite{Rasouli:2022tjn, Rasouli:2023mae}; the Greek indices run from zero to $n-1$; $\phi$ is a scalar field minimally coupled to the Ricci scalar $R^{^{(n)}}$; $\nabla$ denotes the covariant derivative; $g$ denotes the determinant of the $n$--dimensional metric $g_{\alpha\beta}$; $J(\phi)$ is a function of the scalar field specifying the non-canonical kinetic structure of the model. We should note that in SB theory, it has been assumed that $1/2 {\cal J}(\phi)=\omega \,\phi^r$, where $r$ and $\omega$ are two parameters of the SB model.

It is worth mentioning that, within specific situations, we can
apply the following transformations to re-express both the kinetic and the potential terms in the action \eqref{SB-action}. This procedure allows us to recover, as a particular case, 
the standard Einstein-scalar field system (ESFS), that is, the well-known gravitational framework in which the scalar field with a canonical kinetic term is minimally coupled to gravity \cite{Yi:2021xhw,Rasouli:2022tjn}.

\begin{itemize}
    \item 
Employing $d\varphi=\sqrt{{1/2\cal J}(\phi)}d\phi$, where 
$ {\cal J}(\phi)>0$ can be used to transform the general form of the original action \eqref{SB-action} into one with the expected canonical kinetic term:
\begin{equation}\label{can-action-1}
{\cal S}_{_{\rm SB}}\rightarrow {\cal S}=\int d^{n}\!x \sqrt{-g}\,
 \Big[\frac{1}{2\kappa_n}R^{^{(n)}}- g^{\alpha\beta}\,({\nabla}_\alpha\varphi)({\nabla}_\beta\varphi)
 -U(\varphi)\Big],
\end{equation}
where $U(\varphi)$ is related to the original potential $V(\phi)$ through the transformation $ U[\varphi(\phi)] = V(\phi)$, preserving the potential structure of the initial theory.

Moreover, by expressing the function ${\cal J}(\phi)$ in terms of potentials and their derivatives, the  action \eqref{can-action-1} can be rewritten as
\begin{equation} \label{can-action-2}
S = \int d^n x \, \sqrt{-g} \left[ R^{^{(n)}} - \left(\frac{dV(\phi)}{d\phi}\frac{dU^{-1}(V(\phi))}{dV(\phi)}\right)^2 g^{\alpha\beta}(\nabla_{\alpha}\phi)(\nabla_{\beta}\phi) - V(\phi) \right],
\end{equation}
where $U^{-1}$ represents the inverse function of $U$.

    
    \item   In the particular case where $r=0$ and $\omega=1/2$, we get the well-known ESFS in n dimensions.
\end{itemize}

In certain parts of this paper, such transformations, especially the second one, will be employed to simplify the model.

The main objective of the present work is to establish an extended version of the scalar field cosmology through a conceptually new approach that, to the best of our knowledge, has not yet been investigated. We include a fractional time-dependent kernel in the action \eqref{SB-action-standard} in arbitrary dimensions as 
\begin{eqnarray}\label{SB-action}
S^{}_{_{\rm TSB}}=\int\xi({t}) {\cal L}_{_{\rm SB}}d{ t},
\end{eqnarray}
where ${\cal L}_{_{\rm SB}}$ is given by \eqref{SB-action-standard}.

We should emphasize that the time-dependent kernel $\xi(t)$ is not treated as an independent dynamical degree of freedom. In this approach, the kernel does not possess a kinetic term nor an associated equation of motion, and it should be regarded as an external function providing possible memory-like or non-local effects in the cosmological dynamics. Accordingly, the present construction should be interpreted as an \textit{effective} cosmological framework rather than a fundamental covariant theory.

In this section and Section \ref{Pert-Dyn}, both the function $J(\phi)$ and the time-dependent kernel $\xi(t)$ are kept completely general to preserve the generality of the formulation. However, in Section \ref{FLRW-SB-Cosmology}--\ref{SecIV}, regarding the fractional SB-FLRW cosmology and its relevant dynamical systems, the corresponding specific functional forms will be employed when required for analytical tractability and for exploring concrete cosmological scenarios.

Varying the action \eqref{SB-action} with respect to the scalar 
field, we can easily obtain an extended Klein-Gordon (KG) equation: 
\begin{eqnarray}\label{SB-eq0}
 {\cal J} {\nabla}^2 \phi + {\cal J}({\nabla}_\mu \ln{\xi})\partial^\mu \phi+\frac{1}{2}{\cal J}_{,\phi}(\partial_\alpha \phi)(\partial^\alpha \phi)-V_{,\phi}=0.
\end{eqnarray}

Moreover, varying the action \eqref{SB-action} with respect to the metric gives:
\begin{eqnarray}\label{SB-eq1}
G^{(n)}_{\mu\nu}=\kappa_n\left[T^{(\phi)}_{\mu\nu}+T^{(\xi)}_{\mu\nu}\right]\equiv \kappa_n\, T^{\rm{(eff)}}_{\mu\nu} ,
\end{eqnarray}
where we assumed the signature of an n-dimensional metric as $(- ++...+)$ and used the definitions of the stress energy tensors associated with the scalar field and the fractional sector as
\begin{eqnarray}\label{SB-eq2}
 T^{(\phi)}_{\mu\nu}\equiv{\cal J}(\phi)\, \,({\nabla}_\mu\phi)({\nabla}_\nu\phi)-g_{\mu\nu}\left[\frac{1}{2}{\cal J}(\phi)({\nabla}_\alpha\phi)({\nabla}^\alpha\phi)+V(\phi)\right],
\end{eqnarray}
\begin{eqnarray}\label{SB-eq3}
 T^{\rm(\xi)}_{\mu\nu}\equiv\frac{1}{\kappa_n \xi} \,({\nabla}_\mu\nabla_\nu \xi-\nabla^2 \xi). 
\end{eqnarray}
It follows from equation \eqref{SB-eq1} together with the Bianchi identities, that the total (effective) energy-momentum tensor,  $T^{\mathrm{(eff)}}_{\mu\nu}$ 
is conserved, namely, $\nabla^\mu T^{\rm{(eff)}}_{\mu\nu} = 0$.
However, the scalar sector and $\xi$-sector are not separately conserved,
which indicates an effective exchange of energy-momentum 
between the two sectors induced by the time-dependent kernel. At the background cosmological level, this exchange manifests primarily as energy transfer, while at the perturbative level it also involves momentum transfer.

It is straightforward to show that the Ricci scalar $R^{(n)}$ is given by 
 \begin{eqnarray}\label{SB-eq4}
R^{(n)}=\frac{2(n-1)}{n-2}\xi^{-1} \nabla^2 \xi-\frac{2 \kappa_n}{n-2} T^{(\phi)}, 
\end{eqnarray}
where 
 \begin{eqnarray}\label{SB-eq4-1}
T^{(\phi)}=(1-\frac{n}{2}){\cal J}(\phi)(\partial \phi)^2-n V(\phi),
\end{eqnarray}
and we used equations \eqref{SB-eq1}, \eqref{SB-eq2} and \eqref{SB-eq3}.

In the next section where we discuss the Bianchi identities and the 
Noether theorems, we use of the equations derived above.

\section{Perturbative dynamics and symmetry identities}
\label{Pert-Dyn}
In Section \ref{SecII}, we introduced an extended SB model in $n$ dimensions, in which, in addition to the scalar potential, a time-dependent kernel is incorporated, leading to modified field equations. In Sections \ref{FLRW-SB-Cosmology}--\ref{SecIV}, by considering the Friedmann-Lema\^{i}tre-Robertson-Walker (FLRW) metric, and specifying the kernel through a fractional parametrization, we derive exact solutions and construct two dynamical systems. 

In order to complement the analysis of the model introduced in the preceding section, we investigate linear perturbations around a homogeneous and isotropic FLRW background. While the background equations already capture the main features of the cosmological evolution, they do not provide sufficient information about the stability of the model or the behavior of fluctuations. For this reason, a perturbative analysis is required to obtain a more complete characterization of the underlying dynamics.

In the present framework, the extended structure is 
introduced through a time-dependent kernel in the action, which leads to additional non-autonomous contributions in the field equations. As a consequence, the perturbation equations inherit this structure, and may exhibit features that are absent in the corresponding standard scalar field cosmology. In particular, the interplay between the scalar field sector and the fractional kernel is expected to influence the evolution of scalar modes in a non-trivial way.

A further motivation for performing this analysis is to verify the consistency of the model with the standard cosmological scenario. In the particular case where $J=1=\xi$, the model is expected to reduce to the conventional minimally coupled ESFS. At the perturbative level, this implies that the corresponding equations should recover the well-known results of canonical scalar field cosmology, providing a non-trivial consistency check of the formulation.

It is also worth noting that the time-dependent kernel can be interpreted as encoding effective memory effects in the cosmological dynamics. Such effects, while introduced at the level of the action, may leave observable imprints in the perturbative sector, thereby motivating their systematic investigation.

However, we emphasize that the primary goal of this section is to establish a general perturbative framework, rather than to carry out a detailed phenomenological analysis. In particular, a complete study of structure formation or observational constraints is beyond the scope of the present work. 

 We only investigate first-order perturbations around a homogeneous and isotropic cosmological background described by the FLRW metric. We derive the corresponding field equations and incorporating a fundamental analysis of the Bianchi identities together with Noether's second theorem. Such considerations are of central importance for any consistent cosmological framework. To the best of our knowledge, such discussions represent the first attempt to investigate these aspects within time-weighted 
 action framework or its associated fractional realizations.
More specifically, the perturbative analysis (for a detailed study of the cosmological perturbation theory, refer to \cite{bardeen1980gauge,mukhanov1992theory,Durrer:1993tti,Durrer:1994nk} and related papers) provides a more complete description of how scalar fluctuations propagate consistently in the presence of the generalized action with $\xi(t)$. 

In this section, we consider the action \eqref{SB-action} with general functions $J(\phi)$ and $\xi(t)$, without specifying their explicit functional forms. The precise dependence of these functions on the scalar field and time, which leads to a particular fractional SB model, will be specified in subsequent sections.

In this general case, the energy density and pressure associated with the scalar field sector and with the $\xi$-sector (which is denoted by the index ``$\xi$'') are given by the following relations
\begin{eqnarray}\label{ro-p-phi-gen}
 \rho_{_{\phi}} &=& \frac{1}{2} {\mathcal J}\,\dot\phi^2 + V(\phi), \hspace{10mm}
p_{_{\phi}} = \frac{1}{2} {\mathcal J}\,\dot\phi^2 - V(\phi),
\end{eqnarray}\\
\begin{eqnarray}
\label{ro-p-fr-gen}
\rho_{_{\xi}} = -\frac{n-1}{\kappa_n}\frac{H\dot\xi}{\xi}, \hspace{12mm}
p_{_{\xi}} = \frac{1}{\kappa_n\xi}\left[\ddot\xi+(n-2)H\dot\xi\right]
\end{eqnarray}
which are obtained by projecting the stress-energy tensors defined in equations~\eqref{SB-eq2} and \eqref{SB-eq3} onto the FLRW background, i.e., $\rho = -\,T^{0}{}_{0}$ and 
$p = T^{i}{}_{i}$ (no sum).

Let us consider scalar perturbations around a spatially flat FRW background as
\begin{equation}\label{Pert-FLRW}
ds^2=-(1+2\Psi)dt^2+a^2(t)(1-2\Phi)\delta_{ij}dx^i dx^j,
\qquad i,j=1,\dots,n-1,
\end{equation}
where $\Phi$ and $\Psi$ denote scalar metric perturbations. Equation \eqref{Pert-FLRW} corresponds to scalar perturbations of the FLRW metric in $n$ dimensions, written in the Newtonian gauge.

At first order, the anisotropic stress vanishes in both the scalar and kernel-induced sectors, so we can consistently set $\Phi=\Psi$.

Moreover, the scalar field is perturbed as
\begin{equation}\label{pert-phi}
\phi(t, \vec{x}) = \bar{\phi}(t) + \delta \phi(t, \vec{x}).
\end{equation}
where $\bar{\phi}(t)$ obeys the background KG equation \eqref{asli3}. 
Furthermore, we take $\xi=\xi(t)$ with no perturbation in Newtonian gauge: $\delta \xi=0$, $\partial_t \xi=\dot{\xi}$ and $\partial_i \xi=0$. For clarity, we emphasize that the time-dependent kernel $\xi(t)$ is treated as an external function and is not perturbed at the linear level. Consequently, we assume that the fractional kernel remains homogeneous, i.e., $\delta \xi = 0$, so that all perturbations arise solely from the scalar and metric sectors.

In what follows, the key equations of our framework are derived at the level of first-order perturbations, followed by a brief discussion of their theoretical and physical implications.

\subsection{First-order Friedmann equations}

We begin by deriving the perturbative gravitational field equations, and then, within the framework of the cosmological model under consideration, we proceed to obtain the associated Friedmann equations.

\begin{description}
    \item[Energy constraint] 
   At first order, the $00$ component of the field equations \eqref{SB-eq1} yields
$\delta G^0_{\ 0} = \kappa_n \delta T^{0\,\rm{(eff)}}_{\ 0}$, where the effective energy density perturbation is defined as
$\delta\rho_{\rm eff}=\delta\rho_\phi+\delta\rho_{_{\rm fr}}$. We have shown that the linearized $00$ component leads to the energy constraint equation as
\begin{equation}\label{pert-00}
\frac{\nabla^2 \Phi}{a^2}-(n-1)H(\dot\Phi+H\Psi)
=\frac{\kappa_n}{2}\,\delta\rho_{_{\rm eff}}=\frac{\kappa_n}{2}\,(\delta\rho_\phi+\delta\rho_{_{\rm fr}}),
\end{equation}
where
\begin{eqnarray}
\label{pert-rho-phi}
  \delta\rho_\phi &=&{\mathcal J}\,\dot{\bar\phi}\,\delta\dot\phi
-{\mathcal J}\,\dot{\bar\phi}^2\,\Psi
+V_{,\phi}\,\delta\phi
+\tfrac12 {\mathcal J}_{,\phi}\,\dot{\bar\phi}^2\,\delta\phi,\\
\label{pert-rho-fr}
\delta\rho_{_{\rm fr}} &=& -\frac{n-1}{\kappa_n}\,\frac{\dot\xi\,\dot\Phi}{\xi}.
\end{eqnarray}
The perturbations $\delta\rho_\phi$ and $\delta\rho_{_{\rm fr}}$ are obtained by linearizing the corresponding background energy densities, see equations \eqref{ro-p-phi-gen} and \eqref{ro-p-fr-gen}. In particular, the scalar sector includes both field and metric contributions through $\delta(\dot\phi^2)=2\dot\phi\,\delta\dot\phi-2\dot\phi^2\Psi$, while the time-kernel contribution follows from perturbing the combination $H\,\dot{\xi}/\xi$, using $\delta H=\dot\Phi+H\Psi$ and $\delta(\dot{\xi})=-\dot{\xi}\Psi$.

\item [Momentum constraint]
The momentum constraint is obtained from the $0i$ component of the linearized field equations. At first order, the scalar field sector contributes through the velocity perturbation term proportional to $\partial_i\delta\phi$, whereas the $\xi$-sector generates an additional contribution proportional to $(\dot{\xi}/\xi)\partial_i\Psi$ due to the time dependence of the kernel:
\begin{equation}
2\partial_i\left(\dot\Phi+H\Psi \right)
=\kappa_n\left(-\mathcal J\,\dot{\bar\phi}\,\partial_i\delta\phi
+\frac{\dot\xi}{\xi}\,\partial_i\Psi\right),
\end{equation}
that can be simplified in Fourier space as
\begin{equation}\label{pert-0i-Four}
2\left(\dot\Phi+H\Psi \right)
=\kappa_n \left(-\,{\mathcal J}\,\dot{\bar\phi}\,\delta\phi
+\frac{\dot\xi}{\xi}\,\Psi\right),
\end{equation}
which makes explicit that both the scalar sector and $\xi$-sector contribute to the effective momentum density.
In comparison with the corresponding standard case, we observe that the second term in the right hand side of equation \eqref{pert-0i-Four} is a new contribution arising from the $\xi$-sector. It should be noted that 
equation \eqref{pert-0i-Four} ensures consistency 
between the modified SB field equations and the perturbed Klein-Gordon 
equation, reflecting the underlying conservation of the total energy-momentum tensor.

\item[Pressure constraint] Let us now focus on the $ij$ 
 components of the linearized field equations, evaluated on the perturbed FLRW background in Newtonian gauge with $\Phi=\Psi$:
\begin{equation}\label{pert-ij}
2\Big[\ddot\Phi+(n-1)H\dot\Phi\Big]
+2\Big[\dot H+(n-2)H^2\Big]\Psi
=\kappa_n\,\delta p_{_{\rm eff}}=\kappa_n\,(\delta p_\phi+\delta p_{_{\xi}}),
\end{equation}
where
\begin{eqnarray}\label{pert-p-phi}
\delta p_\phi &=&
{\mathcal J}\,\dot{\bar\phi}\,\delta\dot\phi
-{\mathcal J}\,\dot{\bar\phi}^2\,\Psi
- V_{,\phi}\,\delta\phi
+\frac{1}{2} {\mathcal J}_{,\phi}\,\dot{\bar\phi}^2\,\delta\phi,\\
\label{pert-p-fr}
\delta p_{_{\xi}} &=& -2p_{_{fr}}\,\Psi
+\frac{1}{\kappa_n\xi}\Big[(2-n)\dot\xi\,\dot\Phi-\dot\xi\,\dot\Psi\Big].
\end{eqnarray}
The perturbation $\delta p_\phi$ is obtained by perturbing the background pressure (see equation \eqref{ro-p-phi-gen}), taking into account the metric perturbations in the kinetic term. Moreover, the pressure perturbation $\delta p_{_{\xi}}$ follows from the perturbation of the spatial components of $T^{\rm(\xi)}_{\mu\nu}$. In the absence of kernel perturbations ($\delta\xi=0$), the result originates entirely from the metric dependence of the covariant derivatives acting on $\xi(t)$, generating contributions proportional to $\dot{\xi}$ and $\ddot{\xi}$.
\item[Shear constraint]
We now consider the traceless part of the spatial $ij$ components of the linearized field equations. At first order, this immediately leads to
\begin{equation}\label{Sh-cons}
\Phi - \Psi = 0.
\end{equation}
This relation reflects the absence of anisotropic stress in the system. While this is expected for the scalar field sector, it is important to verify that the kernel-induced contribution does not alter this result. The effective energy-momentum tensor associated with the $\xi$-sector is constructed from covariant derivatives of $\xi(t)$, schematically of the form $ T^{\rm(fr)}_{ij} \sim \nabla_i \nabla_j \xi - g_{ij}\Box \xi$.
Since the kernel depends only on time and we work with $\delta\xi=0$, all first-order contributions arising from these terms are proportional to $\delta_{ij}$. Consequently, no traceless component is generated by the $\xi$-sector. 
It follows that the condition $\Phi=\Psi$ does not rely on any specific choice of the kernel function, but only on its homogeneous nature. Furthermore, because the traceless part of the field equations does not introduce any scale-dependent terms, this result holds for all Fourier modes, except for the trivial homogeneous mode.

\end{description}

We emphasize that the  result \eqref{Sh-cons} holds at linear order. At higher orders, non-linear couplings may in general generate effective anisotropic stress, as is well known in standard cosmological perturbation theory.

\subsection{Perturbed Klein-Gordon equation in the presence of the time-dependent kernel}
\label{Pert-KG}
Let us first obtain the background KG equation with the homogeneous background $\phi=\bar\phi(t)$: 
\begin{equation}
\ddot{\bar\phi}
+\Big[(n-1)H+\frac{\dot\xi}{\xi}\Big]\dot{\bar\phi}
+\frac{1}{2}\frac{{\mathcal J}_{,\phi}}{\mathcal J}\,\dot{\bar\phi}^2
-\frac{V_{,\phi}}{\mathcal J}=0, \label{eq:BG-KG}
\end{equation}
where we used \eqref{SB-eq0}.
At first order, we obtain the perturbed KG equation as
\begin{eqnarray} \label{pert-KG}
\delta\ddot\phi
&+&\Big[(n-1)H+\frac{{\mathcal J}_{,\phi}}{\mathcal J}\dot{\bar\phi}+\frac{\dot\xi}{\xi}\Big]\delta\dot\phi
-\frac{\nabla^2}{a^2}\delta\phi
+M_{\rm eff}^2(t)\,\delta\phi \nonumber\\
&=& 2\Psi\,\ddot{\bar\phi}
+\dot{\bar\phi}\Big[\dot\Psi+(n-1)\dot\Phi+2(n-1)H\Psi\Big]
+2\frac{\dot\xi}{\xi}\Psi\dot{\bar\phi}
+\frac{\mathcal J_{,\phi}}{\mathcal J}\Psi\dot{\bar\phi}^2,
\end{eqnarray}
where effective mass is:
\begin{equation}\label{eff-mass}
M_{\rm eff}^2 = \frac{V_{,\phi\phi}}{\mathcal J}
+\frac{{\mathcal J}_{,\phi}}{\mathcal J^2}V_{,\phi}
+\frac{1}{2\mathcal J}\left({\mathcal J}_{,\phi\phi}
-\frac{{\mathcal J}_{,\phi}^2}{{\mathcal J}}\right)\dot{\bar\phi}^2.
\end{equation}

We now discuss several features of the perturbed KG equation \eqref{pert-KG}.
(i) The source terms on the right-hand side arise from the coupling between the scalar field perturbations and the metric fluctuations, as dictated by the 
covariant structure of the KG equation. (ii) The presence of the non-trivial function $J(\phi)$ modifies both the kinetic structure and the effective mass of the perturbations, leading to deviations from the canonical scalar field model.  (iii) The term proportional to $\dot{\xi}/\xi$ acts as an additional friction contribution, modifying the damping of scalar field perturbations compared to the standard case. (iv) In comparison to the standard case, the presence of the time-dependent kernel $\xi(t)$ modifies both the friction term and the coupling to metric perturbations. (v) The effective mass $M_{\rm eff}^2$ given in equation \eqref{eff-mass} incorporates contributions from the scalar potential and the function $J(\phi)$. (vi) As a consistency check, the perturbed KG equation \eqref{pert-KG} can be recovered by taking the time derivative of the energy constraint \eqref{pert-00} and using the momentum and pressure equations \eqref{pert-0i-Four} and \eqref{pert-ij} to eliminate the metric perturbations. This reflects the covariant conservation of the total energy-momentum tensor and confirms the internal consistency of the perturbation system. (vii) In the limit $J=1$ and $\xi=\text{const}$, equation\eqref{pert-KG} reduces to the standard perturbed KG equation \cite{mukhanov1992theory}.
     
\subsection{Generalized Mukhanov-Sasaki Equation}

To derive the extended version of the Mukhanov-Sasaki equation, we combine the perturbed gravitational equations with the perturbed KG equation. In particular, we first use the constraint equations \eqref{pert-00} and \eqref{pert-0i-Four} to eliminate the metric perturbations $\Phi$ and $\Psi$ in terms of the scalar field perturbation $\delta\phi$ and its time derivative. Substituting these expressions into the perturbed KG equation \eqref{pert-KG}, we obtain a closed evolution equation for $\delta\phi$.

Next, by introducing the comoving curvature perturbation $\mathcal{R}$  defined as 
\begin{equation}\label{q_n}
{\mathcal R}\equiv \Phi - \frac{H}{\rho_{_{eff}}+p_{_{eff}}}\,q_{_{eff}},
\qquad q_{_{eff}} \equiv -{\mathcal J}\,\dot{\bar\phi}\,\delta\phi
+\frac{\dot\xi}{\kappa_n\xi}\,\Psi,
\end{equation}
and expressing it in terms of the fundamental perturbation variables, the dynamics can be recast as a second--order differential equation for $\mathcal{R}$. In this process, the effects of the $\xi$-sector naturally appear through 
the effective combination $H + {\dot{\xi}}/{(2\xi)}$, which enters the dynamical coefficients of the equation.

Finally, by defining the canonical variable 
\begin{equation} \label{MS-var}
v \equiv z{\mathcal R},\qquad z^2 \equiv a^2 Q_s,
\end{equation}
where
\begin{equation}\label{Qs}
Q_s \equiv\frac{{\mathcal J}\,\dot{\bar\phi}^2+\frac{3}{2}\frac{\dot\xi^2}{\kappa_n\xi}}
{\left(H+\frac{\dot\xi}{2\xi}\right)^2},
\end{equation}
the resulting equation takes the standard Mukhanov-Sasaki form:
\begin{equation}\label{FMS-eq}
v''+\left(k^2-\frac{z''}{z}\right)v=0, \qquad (c_s^2=1),
\end{equation}
with all modifications encoded in the background-dependent function $z$.

In the particular case where $\dot\xi=0$ and ${\mathcal J}=1$, we obtain the standard single-field result $z=a\dot\phi/H$, and therefore equation \eqref{FMS-eq} reduces to its standard counterpart.

\subsection{Bianchi Identities and Noether Theorems}

According to the the Noether's second theorem, diffeomorphism invariance implies the identity
\begin{equation}\label{Bianchi-Id1}
\nabla^\mu G_{\mu\nu} \equiv 0 \quad \Rightarrow \quad \nabla^\mu T^{\text{eff}}_{\mu\nu}=\nabla^\mu\left[T_{\mu\nu}^{(\phi)}
+T_{\mu\nu}^{(\xi)}\right] = 0.
\end{equation}
Thus, the effective EMT is conserved. Moreover, we obtain an exchange law as
\begin{equation}\label{Ex-EMT}
\nabla^\mu\left[\xi\,T^{(\phi)}_{\mu\nu}\right]=-\frac{1}{2\kappa_n}R^{(n)}\nabla_\nu\xi,
\end{equation}
where $R^{(n)}$ denotes the Ricci scalar. 
Equation \eqref{Ex-EMT} implies explicitly that the scalar field exchanges energy-momentum with the kernel contribution whenever $\dot\xi\neq0$. 
When $\xi={\rm constant}$, the exchange equation reduces to the 
standard conservation law, meaning that the energy-momentum of the 
scalar sector is conserved independently. In the general case, however, there 
is a continuous transfer of energy-momentum between the scalar sector
and $\xi$-sector. This exchange is controlled not only by 
the curvature of spacetime but also by the gradient of the time-dependent 
kernel. More precisely, whenever both $\xi(t)$ and the spacetime 
curvature evolve simultaneously, the scalar field cannot preserve 
its energy-momentum independently but instead exchanges 
it with the geometric sector induced by the time-dependent kernel.

For a homogeneous FLRW background, the time component $\nu=0$ of the exchange equation reads
\begin{equation}
\dot{\rho}_\phi + (n-1)H(\rho_\phi+p_\phi) 
= -\frac{1}{2\kappa_n}\,R^{(n)}\,\frac{\dot{\xi}}{\xi},
\end{equation}
which explicitly shows how the rate of energy transfer in 
the scalar sector is governed by the time variation of $\xi$ and the Ricci scalar curvature.

Within the background dynamics, we have shown that the result 
of the above identity can be represented in the following form
\begin{equation}\label{cons-eff-EMT}
\dot\rho_{_{ eff}}+(n-1)H(\rho_{_{ eff}}+p_{_{ eff}})=0.
\end{equation}
It is straightforward to show that the linearized conservation equations in Newtonian gauge can be represented as
\begin{eqnarray}
\delta\dot\rho_{_{ eff}}+(n-1)H(\delta\rho_{_{ eff}}+\delta p_{_{ eff}})
-(n-1)(\rho_{_{ eff}}+p_{_{ eff}})\dot\Phi
-\frac{\nabla^2}{a^2}q_{_{ eff}}&=0,\\
\dot q_{_{ eff}}+(n-1)Hq_{_{ eff}}+(\rho_{_{ eff}}+p_{_{ eff}})\Psi+\delta p_{_{ eff}}&=0,
\end{eqnarray}
where $q_{_{ eff}}$ is given by \eqref{q_n}.

Finally, let us apply the Noether's first theorem to internal symmetries of the scalar field sector. In the special case where ${\mathcal J}_{,\phi}=0$ and $V_{,\phi}=0$, the action admits a shift symmetry $\phi\to\phi+\epsilon$, leading to the conserved current
\begin{equation}\label{cons-charge}
j^\mu=\xi\,{\mathcal J}\,\partial^\mu\phi,\qquad \nabla_\mu j^\mu=0.
\end{equation}
On the FLRW background this indicates that the comoving charge $a^{n-1}\xi\,{\mathcal J}\,\dot{\bar\phi}$ is conserved.

In summary, in this section, we have shown that the perturbed Friedmann and  KG equations, and the Mukhanov-Sasaki equation are not introduced in an \textit{ad hoc} manner but instead arise coherently from the underlying symmetry structure of our extended model.
In particular, we have shown that the momentum constraint $0i$ acquires a new contribution proportional to $\dot{\xi}/\xi$, which clearly distinguishes our framework from the its corresponding standard scenario. Furthermore, we found that the perturbed Einstein equations are not independent; rather, they are connected as a direct consequence of the diffeomorphism invariance of the action.
These results guaranty that the effective EMT remains covariantly conserved even at the perturbative level. Consequently, including such discussions reinforce the internal consistency of our analysis and makes the derivation of the Mukhanov-Sasaki equation transparent to the reader.

It is worth emphasizing that all modified equations derived in this section consistently reduce to those of the SB model, provided that $J(\phi)$ and $\xi(t)$ are chosen as $J=2\omega\,\phi^r$ and $\xi=1$, where $\omega$ and $r$ are constant parameters of the SB model. In this case, the general structure of the equations becomes compatible with that framework.

However, to the best of our knowledge, a complete perturbative formulation of the SB model and those investigated in the current subsection have not been explicitly developed yet. For this reason, a more direct and reliable comparison can be made with the standard minimally coupled scalar field in general relativity. In this limit, by setting $
J=1=\xi$,
all kernel-induced contributions vanish, and it can be readily verified that the generalized perturbation equations obtained here, including the energy and momentum constraints as well as the generalized Mukhanov-Sasaki equation, reduce exactly to their standard counterparts.
In particular, this reduction ensures that no additional propagating degrees of freedom or unphysical behavior arise in the standard limit.

In forthcoming studies, within the framework of the present model, which has been applied to the inflationary epoch of the early universe \cite{oliveira2026power}, we will provide a more comprehensive analysis of the equations derived in this section.

\section{Fractional S\'{a}ez-Ballester cosmology: FLRW background and analytical solutions}
\label{FLRW-SB-Cosmology}

In this section, we consider an action based on that introduced in the previous section (see equation \eqref{SB-action}),
\begin{eqnarray}\label{SB-fr-action-gen}
 S^{(\alpha)}_{_{\rm SB}}=\frac{1}{\Gamma(\alpha)}\int_0^{\bar{t}}\left(\frac{\bar t - t'}{t_*}\right)^{\alpha-1}{\cal L}_{_{\rm SB}}\,dt',
\end{eqnarray}
where ${\cal L}_{_{\rm SB}}$ is given by \eqref{SB-action-standard}, $\Gamma(\alpha)$ denotes the standard Gamma function: $\Gamma(\alpha)=\int_0^{\infty} t^{x-1} e^{-t} dt$ where $\alpha>0$ is the fractional  parameter and $ t_*$ is a fixed reference time scale introduced solely to render the kernel $\xi$ dimensionless. The functions $\xi$ and $J(\phi)$ are assigned specific functional forms, namely $\xi\equiv\left[{({\bar t} - t'})/{t_*}
\right]^{\alpha-1}$ and $J=2 \omega \phi^r$. The former can be inspired by, and can be interpreted as a simplified implementation of the idea of replacing the standard integration measure with the non-trivial Stieltjes one \cite{calcagni2010quantum,calcagni2010fractal,Calcagni2017}, while the latter corresponds to a particular choice within the SB model \cite{saez1986simple}. This specific construction will be referred to as the fractional SB model. Although some qualitative cosmological features may resemble those obtained in related weighted-measure frameworks such as Refs.~\cite{calcagni2010quantum}, the present construction relies on a different effective implementation of the kernel structure and scalar-field sector. Therefore, the corresponding exact solutions and dynamical properties need not coincide in general.

Let us now further elaborate on the motivation for considering actions of the type \eqref{SB-fr-action-gen}. The model presented in this work can be viewed as a simplified and reduced realization of the framework proposed in \cite{calcagni2010quantum,calcagni2010fractal}. In that construction, the standard integration measure $d^D x$
is replaced by a generalized Stieltjes-type measure, $d\varrho(x) = d^D x\, v(x)$,
where the weight function $v(x)$ encodes deviations from a uniform spacetime structure \cite{calcagni2010quantum}. In order to reproduce anomalous scaling properties, this function is typically chosen in the form of a power law, $v(x) \sim |x|^{D(\alpha-1)}$,
so that the effective spacetime volume scales as $\int d\varrho(x) \sim R^{D\alpha}$,
leading to a Hausdorff dimension $d_H = D\alpha$ \cite{calcagni2010quantum}. This construction is naturally compatible with the picture of a fractal spacetime, where geometric properties depend on the observation scale. On the other hand, the appearance of power-law kernels and nonlocal structures is closely related to fractional calculus, where kernels of the form $(\bar t - t')^{\alpha-1}$
arise and encode memory effects in the dynamics. Altogether, these elements provide a consistent motivation for considering weighted and generalized actions of the type adopted in the present work.

To make terminology precise, let us clarify the meaning of the term
\textit{fractional} used in this work. In the present framework the fractional
structure is introduced through a time-dependent kernel in the
action, rather than through the use of fractional derivatives. This
approach belongs to the class of models commonly referred to as
\textit{fractional action functional}, see, for instance \cite{Shchigolev:2010vh,shchigolev2013fractional,Micolta-Riascos:2023mqo} and references therein. In particular, the weighted action can be viewed as a simplified realization of a  left-sided Riemann--Liouville fractional
integral, where the power--law kernel introduces effective memory--like
features in the cosmological dynamics. Consequently, the variational
procedure is performed using standard calculus with respect to the
dynamical variables, while the fractional effects appear through the
explicit time dependence of the kernel in the action.

In particular, in the Riemann--Liouville formulation of fractional integrals, kernels of the form $\left[{({\bar t} - t'})/{t_*}
\right]^{\alpha-1}$ naturally arise as a generalization of repeated integer-order integration to non-integer order \cite{hilfer2000applications}. These kernels encode memory effects by assigning a time-dependent weight to past contributions, thereby introducing non--locality into the dynamics. Such a power--law kernel has also been employed in other gravitational context including fractional extension of the Newtonian gravity, where it leads to non--trivial modifications of the dynamical behavior \cite{Rasouli:2026uax,Rasouli:2026hfy,Rasouli:2026zyq}
It is worth noting that, although alternative kernel choices (e.g., exponential or finite-memory kernels) are possible and commonly used in other physical contexts, the power-law form plays a distinguished role in fractional frameworks due to its scale-invariant nature and direct connection with the underlying formalism.

Moreover, in the present work, we have investigated the simplest fractional 
extension of the SB model. Nevertheless, this framework can be further generalized by incorporating fractional derivatives such as the Riemann-Liouville, Caputo, or Riesz formulations \cite{calcagni2021classical,Rasouli:2021lgy,shchigolev2021fractional,Varao:2024eig,el2023paradigm,Jalalzadeh:2025uuv,canedo2025quantum}, or even by adopting more fundamental approaches based on developments in fractional calculus and quantum gravity paradigms \cite{calcagni2010fractal,calcagni2010quantum,calcagni2011gravity,calcagni2011discrete,Calcagni2011,calcagni2012geometry,arzano2011fractional,calcagni2012diffusion,calcagni2019multifractional,briscese2026fractal}. 

We recall that the kernel $\xi(t)$ is introduced as an effective time-dependent function rather than a dynamical field, and therefore does not possess an independent equation of motion.

It should be emphasized that the kernel in equation \eqref{SB-fr-action} depends only on the time difference
$\Delta t \equiv \bar t - t'$, rather than on $\bar t$ and $t'$, separately, as \textit{the preferred time} manifestation. In particular, the action is defined over the interval $0 \leq t' \leq \bar t$, where $\bar t$ denotes the evaluation time. This implies that the kernel has a retarded support, in the sense that it receives contributions only from past times $t' < \bar t$ and not from future evolution. Therefore, the structure introduced by the fractional kernel is consistent with causality, as the dynamics at time $\bar t$ depends exclusively on the past history of the system.

With this interpretation in place, we now turn to the cosmological implementation of the model and consider an $n$-dimensional Friedmann-Lema\^{i}tre-Robertson-Walker (FLRW) metric:
\begin{eqnarray}
\label{FRW-met-1}
ds^{2}=-N^2(t) dt^{2}+a^{2}(t)\left(\frac{dr^2}{1-\mathcal K r^2}+r^2d\Omega_{_{n-2}}^2\right),
\end{eqnarray}
where $\mathcal K=-1,0,1$; $N(t)$ and $a(t)$ represent the lapse function and the scale factor, respectively; $d\Omega_{_{n-2}}^2=d\theta_1^2+sin^2\theta_1d\theta_2^2+...+sin^2\theta_1
...sin^2\theta_{n-3}d\theta_{n-2}^2$ for $n\geq3$.

Substituting the Ricci scalar associated with the FLRW metric \eqref{FRW-met-1},
\begin{equation}\label{FRW-Ricci}
R^{^{(n)}}=\frac{2(n-1)}{N^2}\left[\frac{\ddot{a}}{a}+\frac{(n-2)}{2}\left(\left(\frac{\dot{a}}{a}\right)^2+\frac{\mathcal K N^2}{a^2}\right) -\left(\frac{\dot{N}}{N}\right)\left(\frac{\dot{a}}{a}\right)\right],
\end{equation}
into the action \eqref{SB-fr-action-gen}, we obtain
\begin{eqnarray}\nonumber
 S^{(\alpha)}_{_{\rm SB}}
&=&\!\frac{1}{\Gamma(\alpha)}\int_{0}^{\bar t}  \frac{a^{n-1}}{N}\left\{\frac{n-1}{\kappa_n}\left[\frac{\ddot{a}}{a}+\frac{(n-2)}{2}\left(\left(\frac{\dot{a}}{a}\right)^2+\frac{\mathcal K N^2}{a^2}\right)-\left(\frac{\dot{N}}{N}\right)\left(\frac{\dot{a}}{a}\right)\right]\right\}
\xi({t'})dt'\nonumber\\\nonumber\\
&+&\!\frac{1}{\Gamma(\alpha)}\int_0^{\bar{t}} \frac{a^{n-1}}{N}\left[\omega \phi^r{\dot{\phi}}^2-{N}^2V(\phi)\right]\xi(t')dt'.
\label{SB-fr-action}
\end{eqnarray}

Employing the Euler-Lagrange equations,
\begin{equation}
\label{EL}
\frac{\partial  L^{(\alpha)}_{_{\rm SB}}}{\partial q_i}-\frac{d}{dt'}\left(\frac{\partial  L^{(\alpha)}_{_{\rm SB}}}{\partial \dot{q}_i}\right)+\frac{d^2}{dt'\,^2}\left(\frac{\partial  L^{(\alpha)}_{_{\rm SB}}}{\partial \ddot{q}_i}\right)=0,
\end{equation}
where $q_i=\{N,a,\phi\}$,
we can easily obtain the equations of motion as
\begin{eqnarray}\label{asli1}
H^{2}+\frac{\mathcal K}{a^2}&+&2\left(\frac{1-\alpha}{n-2}\right)\left(\frac{H}{t}\right)=\frac{2\,\kappa_n \,\rho_{_{\phi}}}{(n-1)(n-2)},\\\nonumber\\
\label{asli2}
(n-2)\frac{\ddot{a}}{a}&+&\frac{(n-2)(n-3)}{2}\left(H^{2}+\frac{\mathcal K}{a^2}\right)\\\nonumber\!&+&\!(n-2)(1-\alpha)\left(\frac{H}{t}\right)+\frac{(1-\alpha)(2-\alpha)}{t^2}=\!
-\kappa_n \,p_{_{\phi}},\\\nonumber\\
\label{asli3}
\ddot{\phi}+(n-1)H\dot{\phi}&+&\frac{r}{2}\left(\frac{\dot{\phi}^{2}}{{\phi}}\right)\!\!+\!\! \frac{1}{2\omega}{\phi}^{-r}\frac{dV(\phi)}{d\phi}+\left(1-\alpha\right)\left(\frac{\dot{\phi}}{t}\right)=0,
\end{eqnarray}
where we assumed $N=1$ and used the transformation $t'-\bar{t}\equiv t$, $H\equiv\dot{a}/a$, $\rho_{_{\phi}}$ and $p_{_{\phi}}$ to represent the 
energy density and pressure of the homogeneous scalar field:
\begin{eqnarray}\label{rho-phi}
\rho_{_{\phi}}&\equiv&{\omega}{\phi}^{r}\dot{\phi}^{2}+V(\phi),\\\nonumber\\
\label{p-phi}
p_{_{\phi}}&\equiv&{\omega}{\phi}^{r}\dot{\phi}^{2}-V(\phi).
\end{eqnarray}

We should note that equation \eqref{asli2} can also be written as
\begin{eqnarray}\nonumber
\frac{\ddot{a}}{a}=\dot{H}+H^2&=&-\frac{\kappa_n}{(n-1)(n-2)}\left[(n-3)\rho_{_{\phi}}+(n-1)p_{_{\phi}}\right]\\\nonumber\\
\label{asli3-1}
&-&\left(\frac{1-\alpha}{n-2}\right)\left(\frac{H}{t}\right)-\frac{(1-\alpha)(2-\alpha)}{(n-2)}\left(\frac{1}{t^2}\right).
\end{eqnarray}
It can be easily shown that the equations \eqref{asli1}--\eqref{asli3} can also be obtained from \eqref{SB-eq0} and \eqref{SB-eq1}.

Moreover, it is seen that including $\xi(t)$ into the SB action results in modified field equations with explicit time-dependent terms proportional to $1/t$ and $1/t^2$. These additional terms vanish in the limit $\alpha = 1$, thereby reducing our fractional model to its standard (non-fractional) counterpart. In this and future investigations, we will show that such an additional degree of freedom can considerably enhance the dynamics and address some of the outstanding problems in the corresponding standard models, while remaining compatible with cosmological observations.

Furthermore, after performing the transformation from the integration variable $t'$ to the cosmic time $t$, the field equations are expressed entirely in terms of $t$, and no independent parameter associated with $\bar t$ appears in the dynamical system.
The fractional corrections manifest themselves only through explicit time-dependent terms such as $1/t$ and $1/t^2$. These terms encode effective memory-like contributions, while preserving a causal (retarded) structure inherited from the action. In particular, no advanced (future-dependent) contributions arise in the field equations.
Therefore, the present formulation does not introduce a fundamental preferred time scale, and remains consistent with causal evolution.

It is important to clarify the role of the fractional parameter $\alpha$. In the present framework, the effective time--dependent kernel is taken in the form $\xi(t')= \left[{({\bar t} - t'})/{t_*}
\right]^{\alpha-1}$, inspired by the structure of fractional integrals in the Riemann--Liouville formulation. Although in many applications the range $0<\alpha<1$ is considered as the simplest non--integer extension, from a mathematical point of view this construction can be extended to general values $\alpha>0$, and such an extension is consistently adopted in this work.
From a dynamical perspective, this choice leads to 
${\dot{\xi}}/{\xi}={(\alpha-1)}/{t}$,
which explicitly enters the field equations. Therefore, as mentioned, the case $\alpha=1$ represents a special limit in which the fractional contribution vanishes and the standard SB model is recovered, while $\alpha\neq1$ leads to non-trivial modifications of the dynamics. We note that the $\alpha=1$ divide and 
the non-conservation of the energy-momentum tensor in weighted-measure cosmological frameworks were originally discussed in \cite{calcagni2010quantum}.
In particular, for $0<\alpha<1$, the kernel is decreasing in time and $\dot{\xi}<0$, resulting in an effective dissipative-like contribution. In contrast, for $\alpha>1$, the kernel grows with time and $\dot{\xi}>0$, so that past contributions are effectively enhanced, leading to a qualitatively different dynamical regime. Hence, the parameter $\alpha$ controls transitions between distinct physical regimes rather than being an arbitrary free parameter.


From now on, let us focus only on the spatially flat FLRW metric where ${\mathcal K}=0$.\\

In the particular case where $\alpha=1$, using relations \eqref{rho-phi}, \eqref{p-phi} and 
the KG equation \eqref{asli3}, it is easy to show that the energy density and pressure obey the conservation law:
\begin{eqnarray}
\label{cons-con}
\dot{\rho}_{_{\phi}}+(n-1)H\left(\rho_{_{\phi}}+p_{_{\phi}}\right)=0.
\end{eqnarray}
Note also that the conservation equation \eqref{cons-con} can be obtained using only the equations \eqref{asli1} and \eqref{asli2}, even without using the relations \eqref{rho-phi}, \eqref{p-phi} and the KG equation.

Let us highlight the following points regarding the field equations of this standard SB Model. The KG equation can be easily derived using the definitions \eqref{rho-phi}, \eqref{p-phi}, and applying the first and second Friedman equations. This, and the aforementioned note regarding the conservation law, may imply that among the four equations \eqref{asli1}-\eqref{asli3} and \eqref{cons-con} only two are independent.
In the following section, we will show how the points mentioned are affected by the fractional model. Thus, different strategies must be considered to solve the equations and analyze the dynamics of these different models.
More concretely, it is correct that all the equations of the standard model are recovered at the level of the field equations (i.e., setting $\alpha=1$), but it is worth noting that there are significant differences between the two frameworks that lead to two entirely different approaches to obtain solutions and analyze their dynamics.


\subsection{The Key Features of the fractional SB model}
\label{SecIII}

In this subsection, let us emphasize several essential aspects of the coupled non-linear differential equations associated with our fractional SB model. 

\begin{itemize}
 \item    
Despite the standard SB model, in the fractional model, the energy 
density and the pressure associated with the scalar field do not satisfy the conservation equation. 
More concretely, we have shown that equation \eqref{cons-con} generalizes to 
\begin{eqnarray}\nonumber
\dot{\rho}_{_{\phi}}&+&(n-1)H\left(\rho_{_{\phi}}+p_{_{\phi}}\right)\\\nonumber\\
&=&-\left(\frac{1-\alpha}{t}\right)\left(2 \omega \phi^{r }\dot{\phi}^2 \right)
=-\left(\frac{1-\alpha} {t}\right) \left(\rho_{_{\phi}}+p_{_{\phi}} \right)\neq0,
\label{cons-con-frac}
\end{eqnarray}
where we have used the relations \eqref{rho-phi}, \eqref{p-phi} and the fractional KG equation.

An important feature emerging from equation \eqref{cons-con-frac} is that the additional time-dependent terms can be interpreted as signaling a departure from the standard local conservation of the energy-momentum tensor. This behavior is in fact consistent with the general structure of models based on generalized integration measures. As already reflected in the other relations of the model (see, in particular, the equation \eqref{Ex-EMT}), the standard conservation law is replaced by a modified relation.
This modification implies that, in the presence of a non-trivial weight function \( v(x) \), the system effectively exchanges energy and momentum with an underlying structure encoded in the measure. Such a feature is fully in line with the framework developed within \cite{calcagni2010quantum}, where the introduction of a Stieltjes-type measure leads to a non-conservative behavior at the local level, while admitting a consistent interpretation in terms of interactions with a nontrivial geometric background. From this perspective, the extra terms appearing in equation \eqref{cons-con-frac} can be understood as an effective manifestation of non-local and scale-dependent effects inherent to this class of models.

A central question concerning our model is whether the Bianchi identities and Noether's second theorem remain fully satisfied in the presence of the fractional sector, which intrinsically involves additional geometrical quantities. At first glance, this structure might seem to blur the conventional distinction between geometry and matter in our framework. This issue has been rigorously investigated in the previous section through the construction of the effective energy-momentum tensor (EMT), both at the background and first-order perturbative levels. 
 Furthermore, in the following section, when the equations are reformulated in terms of the effective energy--momentum components, this issue will be re-investigated.

 \item 
Unlike the standard case, it is important to emphasize 
that, in the fractional SB model, one cannot derive either the extended continuity equation \eqref{cons-con-frac} or the KG equation \eqref{asli3} solely from the Friedmann equations.
It should be noted that the four fractional equations \eqref{asli1}, \eqref{asli2}, \eqref{asli3}, and \eqref{cons-con-frac}, together form a self-consistent system.

 \item 
In the standard SB model, only two of the coupled differential equations are independent. Therefore, to determine the three unknowns, one additional relation must be imposed. In most cases, the potential is assumed as an explicit function of the scalar field. In contrast, the fractional SB model contains three independent equations for three unknowns, eliminating the need for any supplementary assumption. However, this does not imply that one cannot adopt a specific potential $V(\phi)$: such an additional choice merely leads to particular solutions whose dynamics can be easily compared with those of the corresponding standard model.
However, it is important to stress that this does not correspond to a standard reconstruction procedure in which the scalar potential is imposed through an external input, such as a prescribed time--dependent Hubble parameter in cosmology \cite{kamenshchik2013reconstruction,chervon2018method}. Instead, once the time--dependent kernel is specified within the action, the scalar potential is determined consistently as part of the coupled dynamical system. Therefore, the resulting dynamics arises from the internal structure of the modified field equations rather than from an externally driven reconstruction scheme.

\item 
We note that the qualitative differences observed in the results can be directly traced back to the sign change of the quantity $\dot{\xi}/\xi \propto (\alpha-1)/t$ around $\alpha=1$, which governs the effective contribution of the fractional sector in the field equations.
\end{itemize}

In the next subsection, we then obtain analytical solutions corresponding to a general potential. In addition, in the subsequent sections, we discuss the gravitational structure and key dynamical features of the proposed fractional framework.

\subsection{Analytical Exact Solutions}
Let us focus on a universe in which the scalar field serves as the dominant form of energy. Based on the key points outlined above, let us consider equations \eqref{asli1}, \eqref{asli2} and \eqref{cons-con-frac} as the three independent equations and proceed to determine the unknowns $H(t)$, $\rho_{_{\phi}}$ and $p_{_{\phi}}$. After some manipulations, we can easily obtain the following
\begin{eqnarray}
\label{H-dot-frac}
\dot{H}=-(n-1)H^2 +\left[3n-4+(2-n)\alpha \right]\left(\frac{H}{t}\right)+\frac{\alpha ^2-3 \alpha +2}{t^2}.
\end{eqnarray}
An exact solution of equation \eqref{H-dot-frac} is:
\begin{eqnarray}
\label{H-frac}
H(t)=\frac{A h(t)-(n-2)\alpha +3 (n-1)}{2 (n-1) t},
\end{eqnarray}
where 
\begin{eqnarray}
\label{h-A}
h (t)\equiv \frac{t^A-C}{t^A+C}, \hspace{10mm} A=A(\alpha,n)\equiv \sqrt{(\alpha -3)^2 n^2+2 (3 \alpha -5) n+1}.
\end{eqnarray}
 In equation \eqref{h-A}, $C$ is an integration constant. 
Since the auxiliary function $h(t)$ is dimensionless, it follows from
\eqref{h-A}
that the integration constant $C$ has dimension $[C]=T^A$, and therefore defines a characteristic time scale of the system. In the asymptotic regimes $C \gg t^A$ and $C \ll t^A$, the Hubble parameter behaves as $H \propto 1/t$, indicating that the power-law structure becomes independent of $A(\alpha,n)$. In these limits, the parameter $\alpha$ appears only in the amplitude, not in the time dependence.
Moreover, note that for $\alpha>0$, and $n\geq3$, we found that the radicand appeared in \eqref{h-A} is strictly positive.

Employing equation \eqref{H-frac} and the 
other equations of the model, we can obtain the other important physical quantities:
\begin{eqnarray}\label{a-frac}
a(t)\!\!&=&\!\!  a_i t^{\frac{-A-\alpha  (n-2)+3 (n-1)}{2 (n-1)}} \left(t^A+C\right)^{\frac{1}{n-1}},\\\nonumber\\
\label{rho-frac} 
\rho_{_{\phi}}(t)&=&\frac{\left[(n-2)A h(t)+(3-\alpha ) n^2-5 n+2\right] 
\left[A h(t)+(3-\alpha ) n-(3-2 \alpha )\right]}{8 (n-1) t^2},\\\nonumber\\\nonumber\\\nonumber
p_{_{\phi}}(t)&=&\frac{A (n-2)} {8 (n-1) t^2}\left[ -A h^2+2 (\alpha -3) n h+6h-4 t \dot{h}\right]\\
&+&\frac{34-24 \alpha-61 n-4(\alpha -9) \alpha n -(\alpha -3)^2 n^3+2 (\alpha -6) (\alpha -3) n^2}{8 (n-1) t^2},
\label{p-frac}\\\nonumber\\\nonumber\\\nonumber
V(t)&=&\frac{A\left\{2 (n-2) t \dot{h}+A  (n-2)h^2-2 (n-1) h [-\alpha +(\alpha -3) n+5]\right\}}{8 (n-1) t^2} \\
&+&\frac{14 \alpha +(\alpha -3)^2 n^3-2 (\alpha -5) (\alpha -3) n^2+2 (\alpha -12) \alpha  n+41 n-20}{8 (n-1) t^2},
\label{V-frac}\\\nonumber\\\nonumber\\
{\omega}{\phi}^{r}\dot{\phi}^{2}&=&-\frac{A (n-2) t \dot{h}+A (\alpha -n+1) h}{4 (n-1) t^2}-\frac{5 \alpha +(\alpha -3) n^2+(\alpha -6) \alpha  n+10 n-7}{4 (n-1) t^2},
\label{phidot-frac}
\end{eqnarray}
where $a_i$ is an integration constant and $ \dot{h}=dh(t)/dt$.\\

In the next section, we will analyze these solutions in detail. 
However, before moving forward, let us highlight a few important points.

\begin{itemize}

\item 
It is worth clarifying the methodological nature of the exact solutions obtained above. 
In deriving these solutions, we have treated the effective quantities $H$, $\rho_\phi$, and $p_\phi$ as unknown functions and solved the cosmological system without imposing a specific scalar potential or fixing particular values of the S\'aez--Ballester parameters $\omega$ and $r$. 
Therefore, within this effective description, the dependence on the non-canonical structure is involved in $\rho_\phi$ and $p_\phi$, and the resulting solutions may overlap with those of the corresponding canonical scalar-field framework. 
This is precisely one of the reasons why the mapping between the SB formulation and the canonical scalar-field description was discussed in Section \ref{SecII}.
However, this correspondence should not be interpreted as a complete equivalence between the underlying frameworks. 
If one instead solves the coupled non-linear field equations directly by specifying a scalar potential, choosing suitable ans\"atze, or exploring different values of the free SB parameters, the non-canonical structure becomes explicit and may lead to broader classes of analytical or numerical solutions.

    \item 
Using analytical or numerical methods, we were unable to show that the fractional potential becomes zero for acceptable parameter values in the problem. This is in contrast to the corresponding standard models, or even some generalized scalar field cosmology models, where solutions with zero potential exist \cite{Rasouli:2018nwi,Rasouli:2022hnp}.

\item
Within our fractional model, for the standard case where $\alpha=1$, it can be easily shown that the solutions satisfy the equations \eqref{asli1}-\eqref{asli3} and \eqref{cons-con-frac}, resulting in a consistent system. However, it is worth noting that obtaining such particular solutions might be nearly impossible from using the corresponding standard scalar field model, there, we would need to guess the corresponding potential and substitute it into the standard field equations (which consist of two independent equations) to derive the solutions. However, such a guess seems highly improbable.
Therefore, we should note that our fractional model can serve as an appropriate framework to generate various potentials (corresponding to different values of $\alpha$) without introducing additional assumptions beyond the chosen effective framework. This approach enables the construction of different scalar field cosmology models where all equations remain consistent.

\item 
We should note that the exact solutions obtained above satisfy all the fractional field equations, especially the fractional KG equation \eqref{asli3}. In this respect, we can rewrite the fractional KG as
\begin{eqnarray}\label{KG-phi-dot}
\left[\frac{d}{dt}+\frac{2}{t^{1-\alpha} a^{n-1}}\frac{d}{dt}\left(t^{1-\alpha} a^{n-1}\right)\right](\omega \phi^r \dot{\phi}^2)+\frac{dV}{dt}=0,
\end{eqnarray}
which we checked its satisfaction by replacing the relations for $a(t)$, $V(t)$ and $(\omega \phi^r \dot{\phi}^2)$ from the relations \eqref{a-frac}, \eqref{V-frac} and \eqref{phidot-frac}. 
\end{itemize}
\section{Dynamical behavior of cosmological quantities for different ranges of the fractional parameter}
\label{behav-alpha}

In the previous section, we derived exact solutions for a generic potential within the fractional SB model. In what follows, we show that the fractional cosmological equations can be rewritten in the form of standard cosmology. This procedure leads to an effective fluid description of the modified dynamics, providing a more transparent interpretation of the evolution of the relevant physical quantities based on the solutions obtained above. Furthermore, as will be shown in the next section, the dynamical system formulation becomes significantly simpler within this standard framework.

With this reformulation, equations \eqref{asli1} and \eqref{asli2} can be cast into the following standard form:
\begin{eqnarray}\label{eff-eq1}
H^{2}&=&\frac{2\kappa_n \rho_{_{\text{eff}}}}{(n-1)(n-2)},\\\nonumber\\
\label{eff-eq2}
(n-2)\frac{\ddot{a}}{a}&+&\frac{(n-2)(n-3)}{2}H^{2}=\!
-\kappa_n p_{_{\text{eff}}},\\\nonumber
\end{eqnarray}
where
\begin{eqnarray}
\label{rho-p--eff}
\rho_{_{\text{eff}}}\equiv \rho_{_{\phi}}+\rho_{_{\rm{fr}}},\hspace{10mm}p_{_{\rm{eff}}}\equiv p_{_{\phi}}+p_{_{\rm{fr}}},
\end{eqnarray}
\begin{eqnarray}
\label{rho-fr}
\rho_{_{\text{fr}}}&\equiv&\frac{(n-1)(\alpha-1)}{\kappa_n}\left(\frac{H}{t}\right),\\\nonumber\\
\label{p-fr}
p_{_{\text{fr}}}&\equiv& \frac{(1-\alpha)}{\kappa_n}\left[(n-2)\left(\frac{H}{t}\right)+\frac{(2-\alpha)}{t^2}\right].
\end{eqnarray}

Substituting $H(t)$ from equation \eqref{H-frac} into equations \eqref{rho-p--eff}, we obtain 
\begin{eqnarray}\label{rho-eff-exact}
\rho_{_{eff}}=\frac{(n-2) \left[A h t-n(\alpha -3)+2 \alpha -3\right]^2}{8 (n-1) t^2},
\end{eqnarray}
\begin{eqnarray}\nonumber
 p_{_{eff}}&=&-\frac{(n-2) \left[A t \dot{h}-A h+  (n-2)\alpha-3 (n-1)\right]}{2 (n-1) t^2}\\
 \label{p-eff-exact}
 &-&\frac{(n-2) \left[A h(t)-\alpha  (n-2)+3 (n-1)\right]^2}{8 (n-1) t^2},
 \end{eqnarray}
where we assumed $\kappa_n=1$.

Moreover, using the relations \eqref{rho-eff-exact} and \eqref{p-eff-exact}, we can easily obtain a relation for the effective equation of state (EoS) parameter:
\begin{eqnarray}\label{w-eff-exact}
w_{_{eff}}(t,\alpha, n, C)\equiv\frac{p_{_{eff}}}{\rho_{_{eff}}}=\frac{1}{n-1}\left[2q-(n-3)\right],
\end{eqnarray}
where $q\equiv-(1+\frac{\dot{H}}{H^2})$ is the deceleration parameter. 
Because of including the time-dependent kernel into the SB action, the standard notion of a conserved energy is modified in the present framework. Accordingly, the continuity-like relation obtained from the field equations should be interpreted in an effective sense, namely as describing an exchange between the scalar field and fractional sectors of the model. Concretely, we still obtain an effective continuity equation as:
\begin{eqnarray}\label{eff-Con}
\dot{\rho}_{_{eff}}+(n-1)H \rho_{_{eff}}\left[w_{_{eff}}+1\right]=0.
\end{eqnarray}
It should be noted that all the expressions obtained above consistently reduce to their standard counterparts in four dimensions.

In the following subsections, we present a numerical investigation of the behavior of various cosmological quantities. However, before turning to the analysis of the dynamical behavior of the relevant quantities and their numerical illustration, we first comment on several important points.
\begin{itemize}

    \item 
As follows from the structure of the time-dependent kernel $\xi(t)\sim t^{\alpha-1}$ and the appearance of the term $\dot{\xi}/\xi \sim (\alpha-1)/t$ in the field equations, a change in the sign of $(\alpha-1)$ leads to a qualitative transition in the dynamics.
In particular, for $0<\alpha<1$, the fractional contribution behaves in an effectively dissipative manner, whereas for $\alpha>1$, it acts as an amplifying contribution, enhancing the effect of past evolution. Therefore, one expects distinct qualitative behaviors in the corresponding cosmological quantities for these two regimes.
Moreover, since the parameter $\alpha$ affects observable cosmological quantities, it can be regarded as a phenomenological parameter that could, in principle, be constrained by observational data. In this sense, exploring different values of $\alpha$ may provide insight into the physically viable range of the model.

\item 
All quantities are computed using a fixed set of initial conditions and parameter values, as specified in the corresponding figure captions. No rescaling has been applied, so the results directly reflect the physical behavior of the system.

\item 
We emphasize that, although the plots correspond to particular choices of parameters, our numerical analysis indicates that the qualitative behavior of the solutions remains robust across a wide range of values. In particular, for moderate values of the integration constant $C$ (i.e., neither extremely large nor extremely small), the overall dynamical features of the system are insensitive to its precise value. Therefore, the results presented below can be regarded as representative of the generic behavior of the model in each regime.

\end{itemize}

\subsection{Case I: $0<\alpha<1$}

In what follows, we analyze the dynamical behavior of the relevant physical quantities in the regime $0<\alpha<1$.
\medskip

\noindent

\begin{itemize}
    \item  \textbf{Bounce behaviour:} 
The left panel of figure~\ref{B.a-H-hdot} shows the evolution of the scale factor $a(t)$, which 
exhibits a non--vanishing minimum at $t_b \simeq 0.705$,
indicating the absence of an initial singularity, $a(t_b)\simeq1.05$. 
The occurrence of a non--singular bounce is clearly demonstrated in figure~\ref{B.a-H-hdot}. We note that similar bounce behavior has also been reported in \cite{calcagni2010quantum,calcagni2013multi}, where weighted actions based on generalized (Stieltjes-type) integration measures were considered.

The right panel of figure~\ref{B.a-H-hdot} displays the Hubble parameter $H(t)$ together with its time derivative $\dot{H}(t)$. One observes that $H(t)$ evolves from negative to positive values, while at $t=t_b$ we have $H(t_b)=0$ and $\dot{H}(t_b)>0$,
which are the standard conditions for a cosmological bounce. Therefore, the model describes a smooth transition from a contracting to an expanding phase.
\begin{figure}
\centering\includegraphics[width=2.9in]{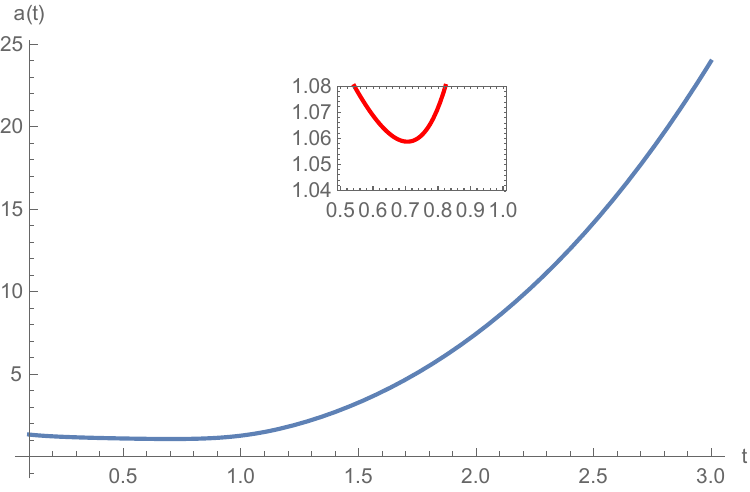}
\centering\includegraphics[width=2.9in]{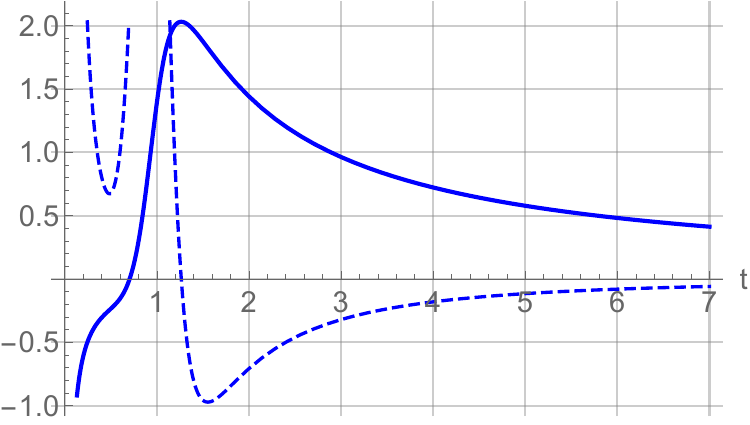}
\caption{Left panel: Evolution of the scale factor $a(t)$ for $0<\alpha<1$, showing a non-vanishing minimum at $t_b\simeq 0.705$, which indicates the occurrence of a non-singular bounce. The inset zooms into the vicinity of the bounce point to clearly illustrate the non--singular minimum of the scale factor.
Right panel: Evolution of the Hubble parameter $H(t)$ (solid line) and its time derivative $\dot{H}(t)$ (dashed line). We assumed $8 \pi G=1$, $\alpha=0.35$, $C=1=a_i$ and $n=4$.
}
\label{B.a-H-hdot}
\end{figure}
\medskip

\noindent
\item\textbf{Pre- and post-bounce dynamics:}
The dynamical nature of the evolution is further 
illustrated in figure~\ref{B.q-weff}, where the deceleration parameter $q(t)$ and the effective equation of state parameter $w_{\rm eff}(t)$ are presented. From figure~\ref{B.q-weff}, we observe that $q(t)<0$ throughout the entire evolution. In particular:
\begin{itemize}
\item For $t<t_b$, the universe is in a contracting phase with $H<0$ and $q<0$, corresponding to an accelerated contraction.
\item For $t>t_b$, the universe enters an expanding phase with $H>0$ and $q<0$, corresponding to an accelerated expansion.
\end{itemize}

\begin{figure}
\centering\includegraphics[width=2.9in]{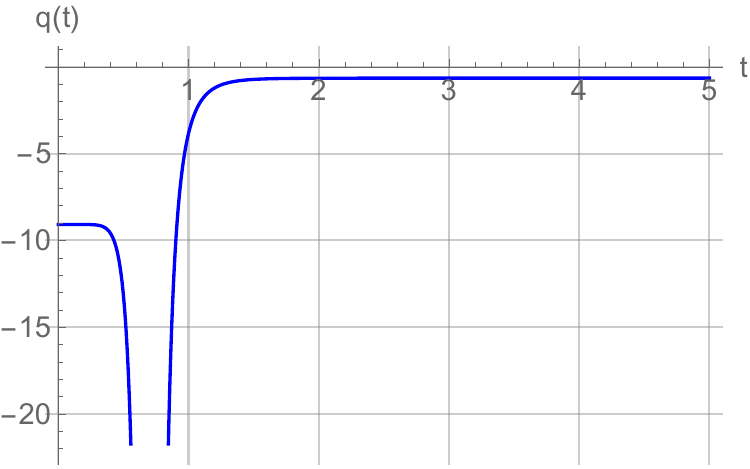}
\centering\includegraphics[width=2.9in]{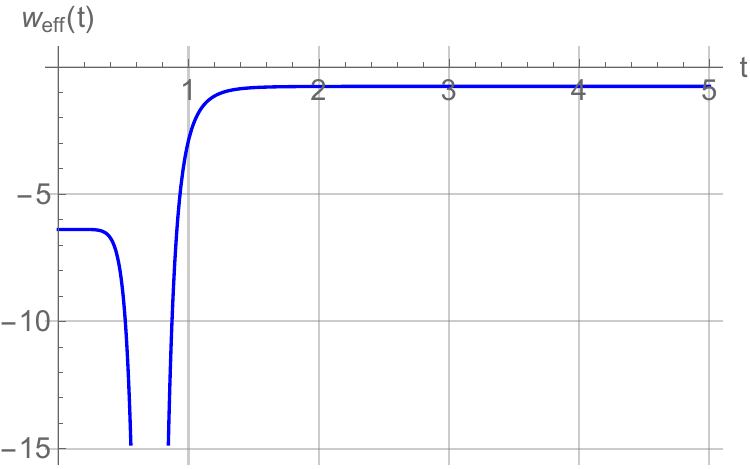}
\caption{Evolution of the deceleration parameter $q(t)$ and the effective equation of state parameter $w_{\rm eff}(t)$ for $0<\alpha<1$. 
Moreover, we assumed $8 \pi G=1$, $\alpha=0.35$, $C=1=a_i$ and $n=4$.
}
\label{B.q-weff}
\end{figure}
\medskip

\noindent
\item\textbf{Energy conditions and origin of the bounce:}
The physical origin of the bounce can be understood from the time behavior of the quantity $\mathcal{N}_{\rm eff}$, which is defined as
\begin{equation}\label{H-dot-N-eff}
\mathcal{N}_{\rm eff} \equiv\rho_{\rm eff} + p_{\rm eff}, \qquad \dot{H} = -\left(\frac{\kappa_n}{n-2} \right)\mathcal{N}_{\rm eff}\, .
\end{equation}
To obtain \eqref{H-dot-N-eff}, we used equations \eqref{eff-eq1}, \eqref{eff-eq2} and \eqref{rho-p--eff}. As figure~\ref{B.Neff} shows, around the bounce point, $\mathcal{N}_{\rm eff}$ becomes negative, indicating a violation of the null energy condition (NEC), which is a necessary ingredient for realizing a non--singular bounce.

Consistently, the effective equation of state shown in figure~\ref{B.q-weff} satisfies $w_{\rm eff} < -1$ in the vicinity of the bounce, corresponding to a phantom--like regime.

\begin{figure}
\centering\includegraphics[width=2.9in]{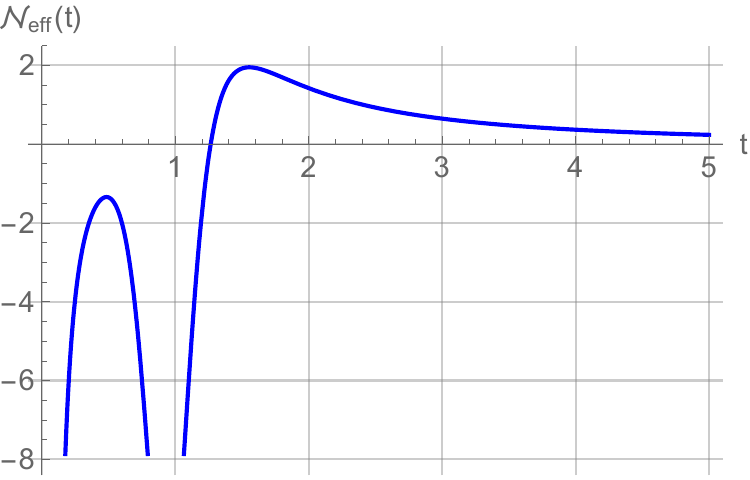}
\centering\includegraphics[width=2.9in]{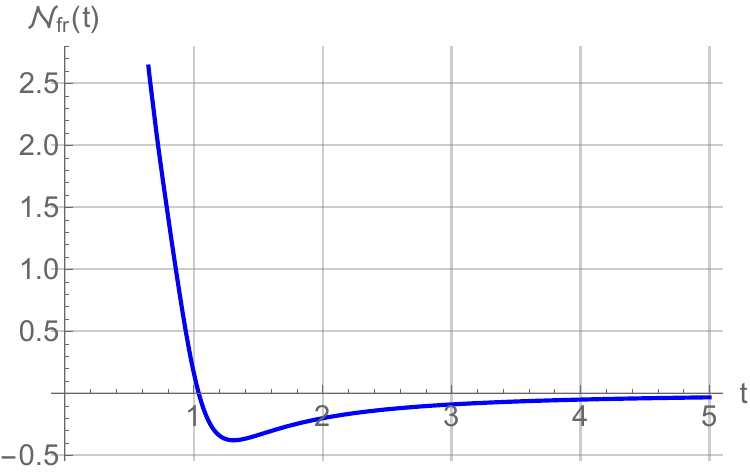}
\caption{Evolution of the effective quantity $\mathcal{N}_{\rm eff}=\rho_{\rm eff}+p_{\rm eff}$ and $\mathcal{N}_{\rm fr}=\rho_{\rm fr}+p_{\rm fr}$ for $0<\alpha<1$. 
We assumed $8 \pi G=1$, $\alpha=0.35$, $C=1=a_i$ and $n=4$.
}
\label{B.Neff}
\end{figure}

\medskip

\noindent
\item\textbf{Role of the scalar and fractional sectors:}
Using relations \eqref{rho-p--eff}, \eqref{H-dot-N-eff} and \eqref{N-fr}, we obtain
\begin{equation}\label{N-fr}
\mathcal{N}_{\rm fr} \equiv \rho_{_{\rm fr}} + p_{_{\rm fr}} = \frac{1-\alpha}{\kappa_n}\left[-\frac{H}{t} + \frac{2-\alpha}{t^2}\right],
\end{equation}
\begin{equation}\label{N-phi}
\mathcal{N}_{\phi} \equiv \rho_{\phi} + p_{\phi} = \mathcal{N}_{\rm eff} -\mathcal{N}_{\rm fr}\,.
\end{equation}
Figure~\ref{B.Neff} also shows that the bounce is primarily driven by the scalar field sector, whereas the fractional modification influences the detailed structure of the evolution.

\medskip

\noindent
\item\textbf{Late-time behaviour:}
At sufficiently late times, the evolution tends toward a regime characterized by a nearly constant positive Hubble parameter, as seen in figure~\ref{B.a-H-hdot}. Correspondingly, the deceleration parameter approaches values close to $q\simeq -1$, while the effective equation of state stabilizes near $w_{\rm eff}\approx -1$ (see figure~\ref{B.q-weff}). 

This indicates that the universe evolves toward a de Sitter--like phase, although no claim of an exact de Sitter solution is made.

\end{itemize}

In summary, for $0<\alpha<1$, the model provides a consistent cosmological scenario characterized by a non--singular bounce, NEC violation around the bounce, a phantom--like effective fluid, and a fully accelerated evolution that asymptotically approaches a de Sitter--like regime. The qualitative features of this behavior remain robust under variations of the integration constant $C$ within a moderate range.

\subsection{Case II: $\alpha>1$}

In this subsection, the analysis of the behavior of the quantities is restricted to the range $1<\alpha<2$. The results are briefly presented below.

\begin{figure}
\centering\includegraphics[width=2.9in]{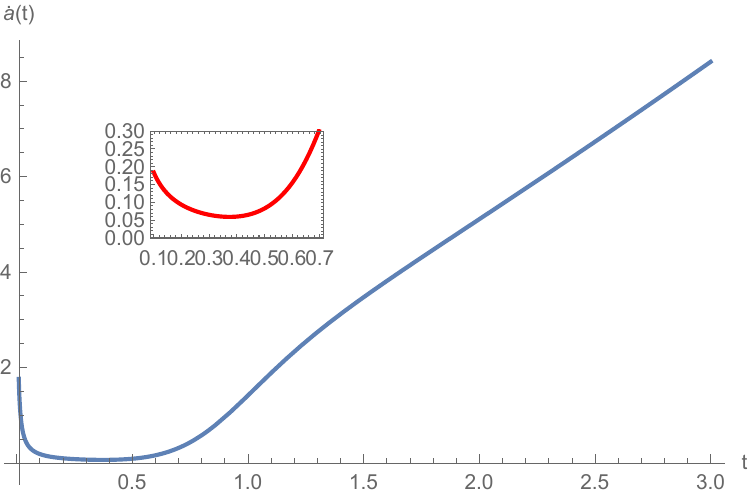}
\centering\includegraphics[width=2.9in]{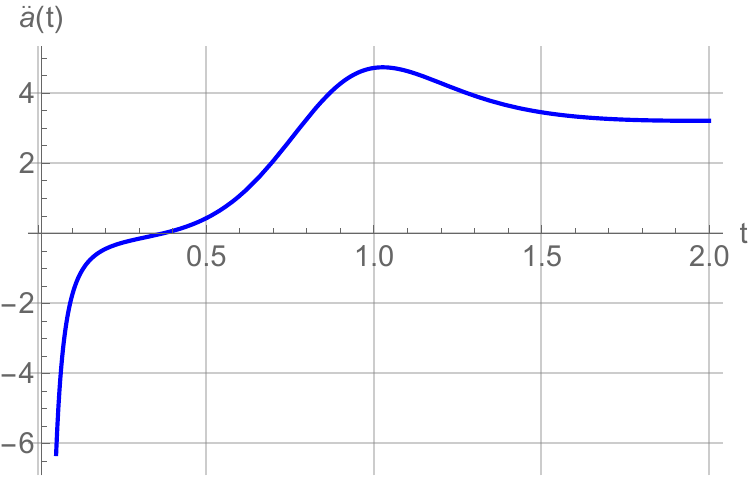}
\caption{Evolution of the first and second derivatives of the scale factor for $\alpha>1$. Left panel: $\dot{a}(t)$; right panel: $\ddot{a}(t)$. The inset in the left panel shows a zoom near the minimum of $\dot{a}(t)$, confirming that it remains strictly positive. 
We assumed $8 \pi G=1$, $\alpha=1.15$, $C=1=a_i$ and $n=4$.
}
\label{II-adot-a2dot}
\end{figure}

\begin{figure}
\centering\includegraphics[width=2.9in]{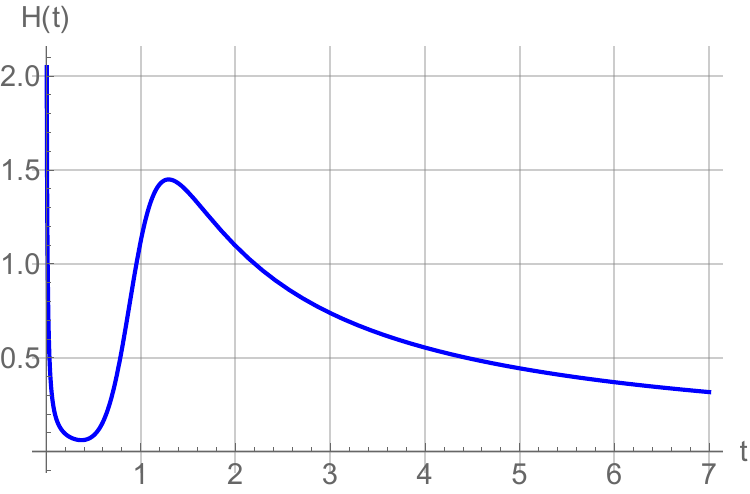}
\centering\includegraphics[width=2.9in]{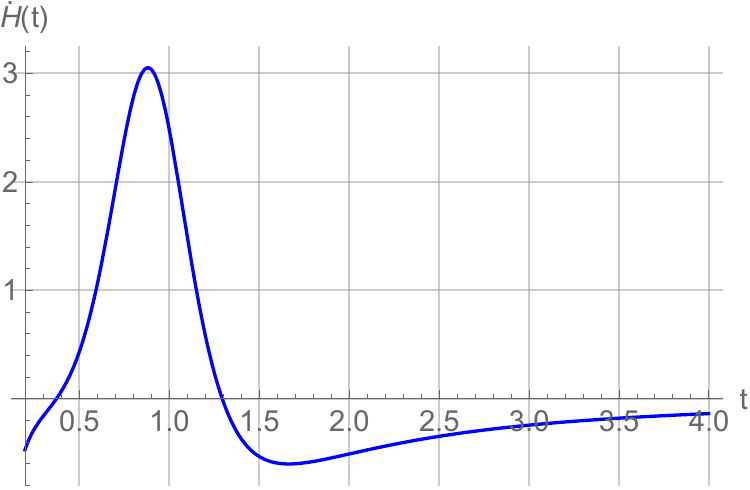}
\caption{Evolution of the Hubble parameter $H(t)$ (left panel) and its time derivative $\dot{H}(t)$  (right panel) for the case $\alpha>1$. We assumed $8 \pi G=1$, $\alpha=1.15$, $C=1=a_i$ and $n=4$. }
\label{II-H-Hdot}
\end{figure}

\begin{figure}
\centering\includegraphics[width=2.9in]{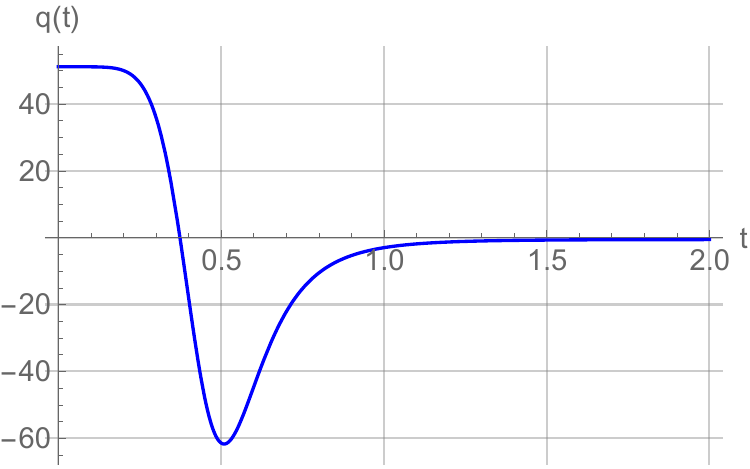}
\centering\includegraphics[width=2.9in]{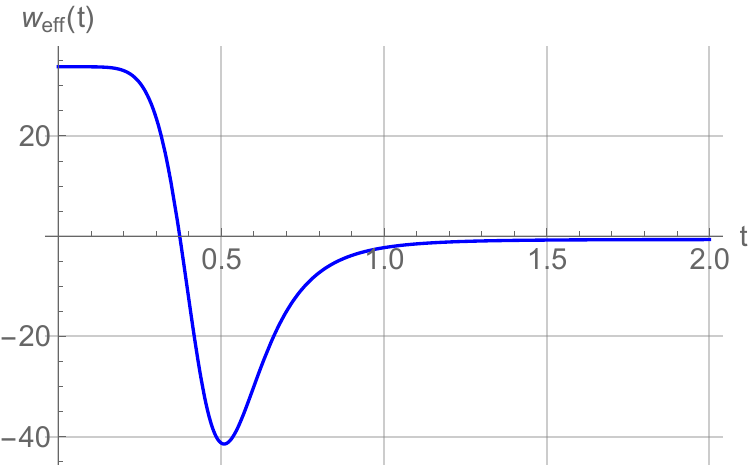}
\caption{Evolution of the deceleration parameter $q(t)$ and the effective equation of state parameter $w_{\rm eff}(t)$ for $\alpha>1$. 
We assumed $8 \pi G=1$, $\alpha=1.15$, $C=1$ and $n=4$.
}
\label{II-q-weff}
\end{figure}

\begin{figure}
\centering\includegraphics[width=3in]{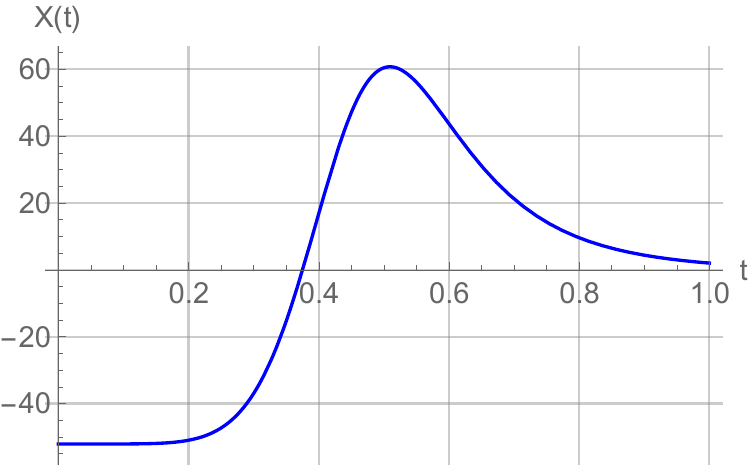}
\centering\includegraphics[width=3in]{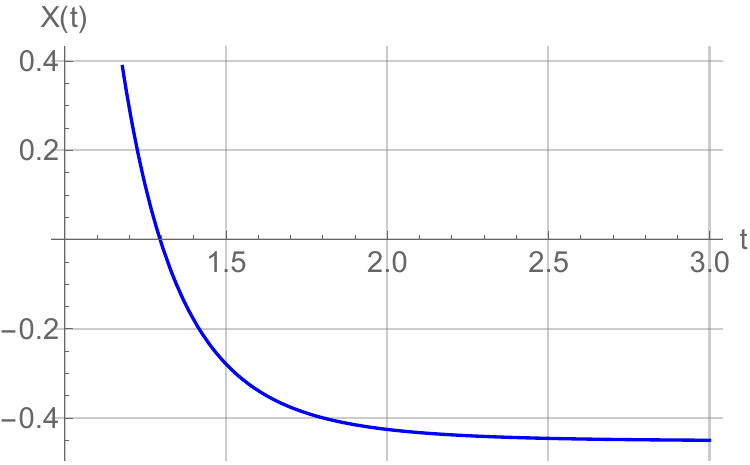}
\caption{Evolution of the dimensionless quantity $X(t)\equiv\dot{H}/H^2$ for $\alpha>1$. This parameter controls both the deceleration parameter $q(t)$ and the effective equation of state $w_{\rm eff}(t)$. For clarity, the early--time behavior is shown in the interval $0 \leq t \leq 1$ (left panel), while the later evolution is displayed for $t \geq 1$ (right panel). We assumed $8\pi G = 1$, $\alpha = 1.15$, $C = 1$, and $n = 4$.
}
\label{II-X}
\end{figure}

\begin{figure}
\centering\includegraphics[width=3in]{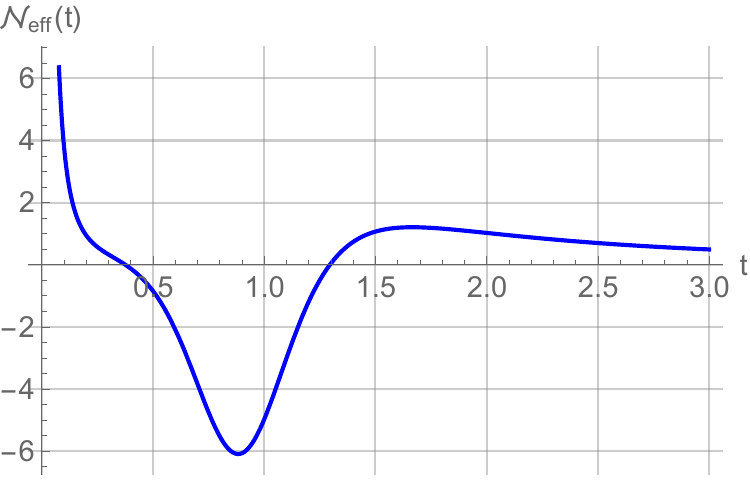}
\centering\includegraphics[width=3in]{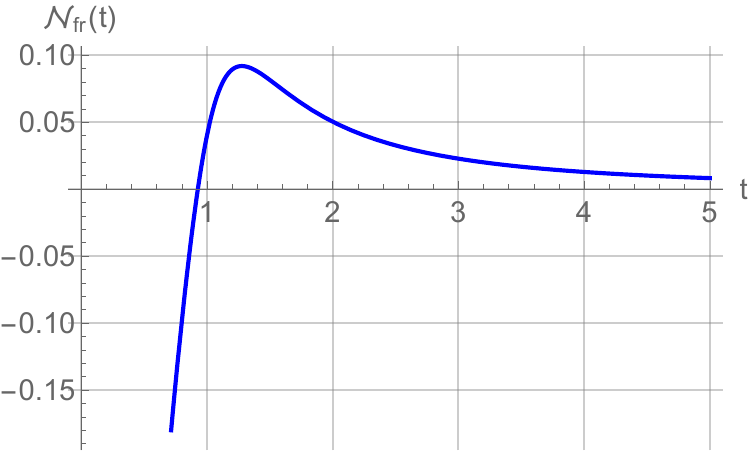}
\caption{Evolution of $\mathcal{N}_{\rm eff}(t)$ (left panel) and $\mathcal{N}_{\rm fr}(t)$ (right panel) for the case II. We assumed $8\pi G = 1$, $\alpha = 1.15$, $C = 1$, and $n = 4$.
}
\label{II-N}
\end{figure}

\begin{itemize}
    
    \item 
Our numerical investigations have shown that the scale factor exhibits a smooth and monotonically increasing behavior throughout the evolution. Although its growth rate changes non-trivially, no genuine minimum is present. This is confirmed by the behavior of $\dot{a}(t)$, which remains strictly positive at all times, indicating that the universe is always 
expanding, see figure \ref{II-adot-a2dot}.
The apparent flattening of $a(t)$ at early times corresponds instead to a minimum in $\dot{a}(t)$, indicating a temporary slowing down of the expansion rather than a contraction phase. This interpretation is further supported by the behavior of $\ddot{a}(t)$, which changes sign from negative to positive, marking a transition from a decelerating to an accelerating expansion phase, see figure \ref{II-adot-a2dot}. Therefore, in contrast to the case~I, the present regime does not exhibit a cosmological bounce, but rather a purely expanding universe with a non-trivial dynamical transition.

\item
The Hubble parameter initially decreases, reaching a non-zero minimum, and then increases before eventually decaying at late times, see figure \ref{II-H-Hdot}. 
Moreover, according to the right panle of the figure \ref{II-H-Hdot}, $\dot{H}$ exhibits a non-trivial structure, changing sign from negative to positive and back to negative. This behavior indicates a sequence of dynamical transitions.

\item
The deceleration parameter $q(t)$ and the effective equation of state parameter $w_{\mathrm{eff}}(t)$ show large variations, including very large positive and negative values at early times, see figure \ref{II-q-weff}. However, these features are not pathological. To clarify their origin, we have also analyzed the dimensionless ratio $X(t) \equiv {\dot{H}}/{H^2}$
which directly controls both $q$ and $w_{\mathrm{eff}}$.

We find that the large values of $q$ and $w_{\mathrm{eff}}$ are associated with regions where $H$ becomes small, leading to large values of $X(t)$. To make this behavior transparent, we have plotted $X(t)$ in two separate time intervals, since its structure is difficult to resolve in a single plot, see figure \ref{II-X}. In regions where $X(t)$ is of order unity, both $q$ and $w_{\mathrm{eff}}$ take physically reasonable values.

\item 
At late times, the system approaches a quasi--asymptotic regime in which both $q$ and $w_{\mathrm{eff}}$ tend to constant values, see figure \ref{II-q-weff}. With the same values of the parameters used to plot the figures in this subsection, our numerical analysis indicates that for large times, we get $q(t) \simeq -0.55$ and $w_{{\rm eff}}(t) \simeq -0.7$, which corresponds to a sustained accelerated expansion, although not exactly of de Sitter type.

\item 
We have also analyzed the time behaviors of the $\mathcal{N}_{\rm eff}$, $\mathcal{N}_{\rm eff}$ and $\mathcal{N}_{\rm eff}$, which confirm the occurrence of the phantom-like phase are driven by the underlying effective sector, see figure \ref{II-N}. Moreover, our investigations show that the fractional contribution modifies the dynamics but does not induce a bounce in this regime.

\end{itemize}

In summary, the case $\alpha>1$ describes a purely expanding cosmology with a sequence of transitions, including a decelerating phase, a transient phantom-like regime, and a late-time accelerated expansion. This behavior clearly demonstrates that the parameter $\alpha$ controls the qualitative cosmological dynamics, interpolating between bouncing and non--bouncing scenarios.

As a final remark of this section, we note that all relevant quantities analyzed in both regimes ($\alpha<1$ and $\alpha>1$), including the Hubble parameter, its time derivative, the deceleration parameter, and the effective equation of state, exhibit the same qualitative behavior for different values of the dimensionality $n$. In this analysis, we have kept all other parameters fixed to the same values used in the previous figures within each subsection, only varying $n$. While quantitative features, such as the position of extrema or the precise values of minima and maxima, may shift slightly, the overall dynamical evolution and physical interpretation remain unchanged in both regimes.

\section{Fractional Dynamical Systems}
\label{SecIV}

In what follows, we construct two possible formulations for the dynamical system of the fractional SB model established in Section \ref{FLRW-SB-Cosmology}.  Moreover, we explain the rationale behind each formulation and clarify why one of them proves more suitable for analyzing the solutions.

\subsection{Generalized Autonomous System and its Challenges}

In the first model, we propose a dynamical system similar to that associated with the corresponding standard scalar field cosmological models based on the Einstein--Hilbert action with a minimally coupled scalar field \cite{ng2001applications,copeland2006dynamics}.
To construct such a dynamical system, let us introduce new variables as
\begin{eqnarray}\label{dyn-1-vars}
x\equiv\frac{\dot{\phi}}{\sqrt{(n-1)(n-2)}}\left(\frac{1}{H}\right), \hspace{5mm} y\equiv\sqrt{\frac{V}{n-1}}\left(\frac{1}{H}\right),
\hspace{5mm} z\equiv t H, \hspace{5mm} \lambda\equiv-\frac{V'(\phi)}{V},  \hspace{5mm} N\equiv \ln{a}.
\end{eqnarray}
After performing some calculations, we have shown that the governing dynamical equations of the model are as follows.
\begin{eqnarray}\label{dyn-1-x}
    x'(N) = -x \left( \frac{\dot{H}}{H^2} \right) - (1-\alpha) \frac{x}{z} - (n-1)x + \frac{(n-1)}{2 \omega \sqrt{(n-1)(n-2)}} \lambda y^2,
    \end{eqnarray}
\begin{eqnarray}\label{dyn-1-y}
     y'(N) =- \frac{\sqrt{(n-1)(n-2)}}{2} \lambda x y - \left( \frac{\dot{H}}{H^2} \right) y,
     \end{eqnarray}
     \begin{eqnarray}\label{dyn-1-z}
    z'(N) = 1 + z \left( \frac{\dot{H}}{H^2} \right),
    \end{eqnarray}
    \begin{eqnarray}\label{dyn-1-lambda}
    \lambda'(N) = \sqrt{(n-1)(n-2)} \left[ \lambda^2 - \frac{V''(\phi)}{V(\phi)}  \right] x,
\end{eqnarray}
where 
\begin{equation}\label{dyn-1-Hdot}
  \frac{\dot{H}}{H^2} = \frac{n-1}{2} -\frac{1-\alpha}{z}-\frac{(1-\alpha)(2-\alpha)}{(n-2)z^2}-\kappa_n (n-1)\left[\omega x^2-\frac{y^2}{n-2}\right].
\end{equation}

Equations \eqref{dyn-1-x}-\eqref{dyn-1-lambda} form an autonomous dynamical system that governs the evolution of our fractional cosmological model. 
As noted above, this system is a generalized formulation of the standard (non--fractional) scalar field model, see, for instance, \cite{ng2001applications,copeland2006dynamics}.

We should note that using this dynamical system to analyze the features of our model is extremely difficult, if not impossible. In standard scalar field cosmology, such dynamical systems can be readily employed, since the scalar potential is specified a priori as a function of the field, $V(\phi)$.
However, in our fractional model, the situation is qualitatively different. The scalar potential is not introduced a priori, but instead is determined from the field equations and obtained as an explicit function of cosmic time, $V(t)$ (see equation \eqref{V-frac}). Due to its highly non-trivial structure, it is generally not possible to express it in closed form as $V(\phi)$.


\subsection{Reformulated System with Effective EMT}

Let us now present the second dynamical system of our framework herein. In this procedure, we employ the fractional cosmological equations presented in the standard form, as given in Section \ref{behav-alpha}.”

Equations \eqref{eff-eq1}, \eqref{eff-eq2} and \eqref{eff-Con} can be written as
\begin{eqnarray}\label{dyn-eff-1}
x'\equiv\frac{dx}{d\eta}&=&-\frac{(n-1)}{2}(1+ w_{_{\rm eff}}) x^2,\\
\label{dyn-eff-2}
y'\equiv\frac{dy}{d\eta}&=&-(n-1)(1+ w_{_{\text{eff}}}) x y,
\end{eqnarray}
  where $x\equiv H/H_0$, $y\equiv \rho_{_{{eff}}}/\rho_c$, 
  $\rho_c\equiv \frac{ (n-1)(n-2)H_0^2}{2 \kappa_n}$,  and $\eta \equiv H_0 t$, in which $H_0$ 
  can be regarded the Hubble constant at present \cite{Planck:2018vyg, Brout:2022vxf, Riess:2024vfa,Pascale:2024qjr, Freedman:2024eph}. 


Inspection of equations~\eqref{rho-fr}-\eqref{p-eff-exact} 
reveals that the effective EoS parameter $w_{_{\text{eff}}}$ explicitly 
depends on $\eta$. Consequently, the associated vector field 
\begin{equation}
\mathbf{F}(x,y,\eta)=\Bigg(-\frac{(n-1)}{2}(1+ w_{_{\rm eff}}) x^2,\; -(n-1)(1+ w_{_{\rm eff}}) x y\Bigg)
\end{equation}
 exhibits an explicit time dependence, implying that the 
system of equations \eqref{dyn-eff-1} and \eqref{dyn-eff-2} is intrinsically non-autonomous.

For the analysis of  non-autonomous dynamical systems, 
three fundamental approaches can be considered as follows \cite{bahamonde2018dynamical}. 

\begin{description}

\item[\textbf{ State augmentation}] 
  
   In this approach, one may introduce an auxiliary variable to parametrize the time dependence of the system, for instance by promoting the  (effective) EoS parameter $w(\eta)$ to an independent variable. By specifying a law of motion for this quantity, such as $w' = \Phi(\eta)$, which can be defined phenomenologically or derived from an underlying physical model, one can analyze an extended dynamical system.
       
 If $w'$ can be expressed as a function of the state variables, $w'=\Phi(x,y)$,
the extended dynamical system becomes autonomous.  
However, if the evolution of $w$ depends explicitly on time, i.e. $w' = \Phi(x,y,\eta)$,
the extended system remains non-autonomous but can still be studied in a higher-dimensional phase space. Typical cosmological realization of this procedure include (i) the Chevallier-Polarski-Linder parameterization, where $w$ is treated as a phenomenological function of the scale factor \cite{scherrer2015mapping}, and (ii) the scalar field closure, in which $w$ is derived from the kinetic and potential terms of a dynamical field \cite{tamanini2014dynamics}.

\item[\textbf{Quasi-autonomous (slowly varying) or piecewise-constant}]
In this approach, it is assumed that $w$ varies 
slowly with time, or, equivalently, that it can be 
treated as a locally constant function within each cosmological epoch. 
Under these assumptions, each regime can be analyzed 
separately as an autonomous dynamical system, provided
that appropriate matching conditions are imposed at 
the transitions between successive phases.

\item[\textbf{Closure by state dependence}] 
In this method, we consider a barotropic-type relation or an effective closure of the form $w_{_{\rm eff}} = W(x,y)$ which can also be applied for our model herein. Therefore, the system \eqref{dyn-eff-1} and \eqref{dyn-eff-2} becomes a closed, autonomous two-dimensional system. In what follows, we present a general analysis that holds for any sufficiently smooth function $W$, without assuming any specific functional form. 

By eliminating $\eta$ from equations \eqref{dyn-eff-1} and \eqref{dyn-eff-2}, we obtain
\begin{equation}\label{dyn-simple}
\frac{d y}{d x} = \frac{-(n-1)(1+w)\,x\,y}{-\tfrac{n-1}{2}(1+w)\,x^2}
= \frac{2y}{x},
\end{equation}
which integrates to $y=\mathbf{C}\,x^{2}$, where $\mathbf{C}$ is 
an integration constant. This result indicates that every 
trajectory of the system \eqref{dyn-eff-1} and \eqref{dyn-eff-2} lies on a parabola of this form.
Physically, along each orbit the effective energy density satisfies $\rho_{_{eff}}\propto H^{2}$, which is fully consistent with a Friedmann-like constraint expressed in these normalized variables.

Let us now investigate the nullclines and the corresponding equilibrium sets of the system. From equations \eqref{dyn-eff-1} and \eqref{dyn-eff-2}, we obtain
\begin{align}\label{dyn-sol1}
x'(\eta) = 0 &\iff x=0 \quad \text{or}\quad 1+W(x,y)=0,\\
y'(\eta) = 0 &\iff x=0 \quad \text{or}\quad y=0 \quad \text{or}\quad 1+W(x,y)=0.
\label{dyn-sol2}
\end{align}
Therefore, we find that:
(i) The vertical line $x=0$ corresponds to a continuum of equilibria (non-hyperbolic);
(ii)  The curve $\Sigma=\{(x,y):\,1+W(x,y)=0\}$ consists entirely of equilibrium points (a de~Sitter-like set), whenever it exists;
(iii) The axis $y=0$ is a $y'$-nullcline, and points on it are equilibria only if simultaneously $x=0$ or $(x,0)\in\Sigma$.

Let us now focus on the direction of motion along the invariant parabolas. 
We consider an invariant $y=\mathbf{C}\, x^2$ with $\mathbf{C}\ge 0$ (as $\rho_{_{\text{eff}}}\ge 0$). Equation \eqref{dyn-eff-1} then implies:
\begin{equation}\label{dyn-sol3}
x' = -A\,\Phi_{\mathbf{C}}(x),\hspace{10mm}
\Phi_{\mathbf{C}}(x)\equiv\bigl[1+W\bigl(x,\mathbf{C}\,x^{2}\bigr)\bigr]x^2, \hspace{10mm} A\equiv \frac{n-1}{2}>0.
\end{equation}
Thus:
\begin{itemize}
\item If $1+W(x,\mathbf{C}x^{2})>0$ (radiation- or matter-like), then $x'<0$: in expansion ($x>0$), $H$ monotonically decreases; in contraction ($x<0$), $H$ becomes more negative (its magnitude increases).
\item If $1+W(x,\mathbf{C}\,x^{2})<0$ (phantom-like), then $x'>0$: trajectories drift toward larger $x$; in expansion $H$ grows (super-acceleration), whereas in contraction $x$ moves toward $0^-$.
\item If the orbit reaches $\Sigma=\{(x,y):\; 1+W(x,y)=0\}$, then $x'=0=y'$ and the motion stops (de Sitter-like equilibrium set).
\end{itemize}
The (dimensionless) time of flight is given by
\begin{equation}\label{dyn-sol4}
\Delta\eta \;=\; -\frac{1}{A}\int_{x_0}^{x}\frac{d \zeta}{\zeta^2\,[1+W(\zeta,\mathbf{C}\zeta^{2})]}\,,
\end{equation}
whose convergence controls whether an orbit reaches $\Sigma$ or a boundary in finite or infinite time. 

Let us now investigate the linearization and stability of the equilibrium sets.
The Jacobian matrix associated with the equations \eqref{dyn-eff-1} and \eqref{dyn-eff-2} reads
\begin{eqnarray}\label{dyn-sol5}
J(x,y)
=
-A
\begin{pmatrix}
2x(1+W)+x^{2}W_{x} && x^{2}W_{y}\\
2y(1+W)+2xy\,W_{x} && 2x(1+W)+2xy\,W_{y}
\end{pmatrix}.
\end{eqnarray}
To analyze stability along the vertical equilibrium line $x=0$, we substitute $x=0$ and obtain
\begin{eqnarray}\label{dyn-sol6}
J(0,y)=
-A
\begin{pmatrix}
0 && 0\\
2y\,[1+W(0,y)] && 0
\end{pmatrix},
\end{eqnarray}
whose determinant and trace both vanish. Hence, the line $x=0$ corresponds to a non-hyperbolic (center-type) continuum of equilibria. Physically, this line represents the bounce or static configuration $H=0$, separating the contracting ($x<0$) and expanding ($x>0$) branches.
It is seen that local drift along this line is governed by the leading non-linear term $x'=-A\bigr[1+W\bigl(0,y\bigr)\bigr]x^2$; the sign of $1+W(0,y)$ determines whether nearby trajectories approach ($1+W<0$) or depart ($1+W>0$) from the $x=0$ branch.

Moreover,let us investigate the transverse stability on the de Sitter-like equilibrium curve $\Sigma$.
On $\Sigma$ (and for $x\neq 0$), the Jacobian matrix \eqref{dyn-sol5} reduces to
\begin{eqnarray}\label{dyn-sol7}
J\big|_{\Sigma}=
-A
\begin{pmatrix}
x^{2}W_{x} && x^{2}W_{y}\\
2xy\,W_{x} &&2xy\,W_{y}
\end{pmatrix},
\end{eqnarray}
where the second row is $2y/x$ times the first, so $\det J\big|_{\Sigma}=0$ and one eigenvalue is always vanishes, corresponding to motion tangent to $\Sigma$. The transverse eigenvalue is
\begin{equation}\label{dyn-sol8}
\lambda_{\perp} \;=\; -\,A\,\Bigl(x^{2}W_{x} + 2\,x\,y\,W_{y}\Bigr)\;\Big|_{\Sigma}.
\end{equation}
Therefore,
(i) if $\lambda_{\perp}<0$, $\Sigma$ is transversely attracting;
(ii) if $\lambda_{\perp}>0$, $\Sigma$ transversely repelling;
(iii) if $\lambda_{\perp}=0$, a higher-order (center/degenerate) analysis is required.
Geometrically, the sign of \eqref{dyn-sol8} coincide with the sign of the directional derivative of $W$ along the phase-flow vector $(x,2y)$.

\end{description}

Let us briefly present the main results 
obtained from the dynamical system analyzed in this subsection.
Even without invoking the third approach, that is, closure by state dependence, we can easily show that $y = \mathbf{C}\,x^{2}$, which indicates that the geometry of phase trajectories is independent of $w_{_{\rm eff}}$. We have shown that the explicit time dependence in $1+w_{_{eff}}$ only affects the rate of motion along those invariant parabolas. In equation \eqref{dyn-sol4}, $w_{_{eff}}$ was replaced by $W(x,\mathbf{C}x^{2}))$, evaluated along the trajectory. Since $\rho_{_{eff}}\ge 0$, we assumed that $\mathbf{C}\ge 0$ and, therefore, only the upper branches of the parabolas $y=\mathbf{C}x^{2}$ are physically relevant. The sign of $x$ distinguishes the expanding ($x>0$) or contracting ($x<0$) phases, while the locus $\Sigma$ plays the role of a generalized de Sitter set. The sign of $1+W$ selects the decelerating or super-accelerating drift along each parabola, such that the transitions between cosmological eras correspond to sign changes of $1+W$, or, equivalently, to the approach toward or departure from $\Sigma$.


\section{Summary and conclusions}
\label{SecV}
In this paper, we first introduce a generalized version scalar field cosmology with a non-canonical kinetic term in $n$ dimensions. Under appropriate transformations, these models can be recast into the form of standard scalar field theories with canonical kinetic terms, namely, standard Einstein-scalar field system, ESFS). 
The SB framework, as a particular case of these models, constitutes the main focus of the present work. In particular, our analysis in Sections \ref{FLRW-SB-Cosmology}-\ref{SecIV}is devoted to this theory.

In Section \ref{SecII}, we again focused on the general framework of scalar field cosmology with a non-canonical kinetic term, without assuming a specific metric. We then introduced a time-dependent kernel into the corresponding action and derived the corresponding field equations.

In Section \ref{Pert-Dyn}, we developed the perturbative framework of the cosmological model established in Section \ref{SecII}, where a generic function $\mathcal{J}(\phi)$ multiplies the kinetic term and the kernel $\xi(t)$ enters as a key element in the field equations. We have shown that the first order Friedmann equations play a crucial role in ensuring the consistent propagation of scalar perturbations and are closely related to the corresponding KG equation and the generalized Mukhanov--Sasaki equation.

Furthermore, we demonstrated that the perturbed Einstein equations are not independent, but rather interconnected as a consequence of the diffeomorphism invariance of the action. This observation is directly linked to the Bianchi identities and Noether's second theorem, and guarantees that the effective EMT, including both the scalar and kernel sectors, remains covariantly conserved even at the perturbative level. As a result, the framework not only renders the derivation of the generalized Mukhanov--Sasaki equation transparent, but also strengthens the internal consistency of our fractional model in cosmological analyses.

It is worth briefly describing the results of Section \ref{Pert-Dyn}, and mention the key differences with the corresponding standard model.
In summary, the perturbation equations of the fractional cosmological model exhibit several fundamental differences compared to their standard counterparts ($\xi=\mathrm{const}$, $\mathcal J=1$): 

(i) The damping term acquires an additional contribution from the fractional sector,  
    $(n-1)H \;\longrightarrow\; (n-1)H + \frac{\dot{\xi}}{\xi}$
    which modifies the effective friction acting on the scalar field and can significantly impact inflationary or post-inflationary dynamics.

    (ii) Regarding the momentum constraint, it should be noted that, in addition to the usual scalar field terms, a new contribution proportional to $\dot{\xi}/\xi$ appears,
    which clearly distinguishes the time weighted model from the corresponding standard scenario.

   (iii) In the fractional pressure perturbations,  
    the kernel sector contributes non--trivial terms of the form
   $\dot{\xi}\,\dot\Phi$ and $\dot{\xi}\,\dot\Psi$, introducing a direct coupling between metric perturbations and the time kernel.

(iv) In the generalized Mukhanov--Sasaki equation,   
 the canonical variable $z^2$ is modified to $z^2 = a^2 Q_s$,
    instead of the standard $z=a\,\dot\phi/H$.  
    This modification affects the scalar power spectrum and the spectral index.
    
(v) The perturbation equations are not introduced \textit{ad hoc}, but arise coherently from the Bianchi identities and Noether's second theorem. This guarantees that the total effective EMT remains covariantly conserved even at the perturbative level.

It is also worth noting the following points with respect to the contents of Section  \ref{Pert-Dyn}. (i) As phenomenological methods to obtain the perturbative dynamics may lead to inconsistencies between the perturbation equations and the background equations. Therefore, in order to derive the perturbation equations, an action--based approach has been applied. (ii) In the presence of the time-dependent kernel in our fractional model, we have shown that the action remains diffeomorphism invariant, and therefore the second Noether theorem can still be applied. In fact, the time-dependent kernel acts as an external bath through which the scalar energy–momentum is exchanged with the $\xi$-sector. Consequently, the Bianchi identities remain geometrically valid. (iii) It is important to emphasize that, despite the presence of the kernel $\xi(t)$, the quadratic action for scalar perturbations retains its canonical form. More precisely, the coefficients of the time and spatial derivative terms remain identical. Consequently, the effective sound speed of the perturbations is unchanged and equal to unity ($c_s^2=1$). In other words, in our model, unlike in frameworks inspired by $k$-essence or Horndeski theories (where the sound speed typically deviates from unity), scalar perturbations still propagate at the speed of light, without exhibiting any anomalous dispersion; see, for instance, \cite{Boiza:2024fmr,deBoe:2025sqv} and references therein.
(iv) We should note that all the perturbative equations obtained in Section \ref{Pert-Dyn} reduce, in the special case $\xi=\mathrm{constant}$ and $\mathcal{J}=1$, to the corresponding equations of the standard ESFS.

In Section \ref{FLRW-SB-Cosmology}, a specific form of the generalized time-weighted action has been considered. In particular, by adopting the FLRW metric in $n$ dimensions, the focus has been placed on SB cosmology. By choosing $J(\phi)$ as the corresponding function in this model and inspired by the Riemann-Liouville integral, a fractional SB model has been formulated.

We have demonstrated that this model is more general and fundamental than its corresponding counterparts. Let us be more precise. In our framework herein, three independent coupling differential equations are simultaneously solved to get three distinct unknowns without requiring specific ad hoc assumptions that typically are taken as phenomenological potentials in the corresponding standard models. Concretely, the potential emerges as a function of the fractional parameter obtained by solving the field equations. Notably, such a possibility can encompass a wide range of analytical/numerical solutions.

After deriving the field equations associated with the general case, we focused solely on the spatially flat universe and managed to obtain exact solutions. We found that all key physical variables used to describe the universe are functions of time, the fractional parameter, the number of dimensions, and only a single integration constant, which can be interpreted a characteristic time scale of
the system.

In the special case where $\alpha=1$, all the modified field equations and the corresponding exact solutions are reduced to their standard counterparts. However, a subtle but significant point arises here: obtaining such a particular solution using the standard model is almost impossible, as it is extremely challenging to intuitively hypothesize such a potential in the corresponding models. 
Another unique feature of this fractional scalar field model is that it does not lead to consistent equations when the potential is zero, which makes it different from other corresponding models. More precisely, our research indicates that for no set of parameter values in the problem, the potential becomes zero. In fact, given the previously mentioned points regarding the number of independent equations and the number of distinct unknowns, this feature is understandable.

In Section \ref{behav-alpha}, it has been shown that rewriting the fractional cosmological equations in terms of an effective standard description has provided a transparent framework for analyzing the behavior of the relevant cosmological quantities. Our numerical investigations have shown that the dynamics of the model can be qualitatively divided into two distinct regimes, depending on the value of the fractional parameter $\alpha$. By adopting a common set of numerical values for the parameters, particularly moderate values (neither very large nor very small) for the integration constant $C$, the following results have been obtained:

\begin{enumerate}

\item For $0<\alpha<1$, the model has described a non-singular bouncing universe. In this regime, the scale factor has exhibited a non-vanishing minimum, and the Hubble parameter has changed sign from negative to positive at the bounce point. In the vicinity of the bounce, the null energy condition (NEC) has been violated. Moreover, the effective fluid has displayed a phantom-like behavior, and the late-time evolution of the universe has approached an accelerated phase close to a de Sitter regime.

\item In contrast, for $\alpha>1$, no genuine bounce has occurred, and the universe has remained in an expanding phase throughout its evolution. Nevertheless, the expansion rate and the effective quantities have exhibited non-trivial structures and distinct dynamical transitions.
\end{enumerate}

It should be emphasized that, in both regimes, the quantities $q$ and $w_{\rm eff}$ have attained large values at early times. However, the analysis of the auxiliary quantity $X(t)=\dot{H}/H^2$ has shown that this behavior has originated from the smallness of $H$, and therefore does not indicate any physical instability. Furthermore, it has been observed that the qualitative features of the results have remained robust under variations of the integration constant $C$ and the spacetime dimensionality $n$. Only quantitative details, such as the positions of extrema or the precise values of minima and maxima, have exhibited minor changes.

Although exact solutions for the model under consideration have been obtained in this work, we have nevertheless placed particular emphasis on constructing the corresponding dynamical system. The reason is that the dynamical system formulation provides insight into the global behavior of the model, offering information beyond that contained in particular exact solutions.

More importantly, the primary goal of this work is not limited to deriving exact solutions, but rather to establish a general framework for a modified cosmological model, including perturbative analysis, dynamical system formulation, and cosmological solutions, and to enable comparisons with the corresponding standard models (such as non-fractional scalar field models with canonical kinetic terms or their four-dimensional counterparts). For this reason, the study of dynamical systems has been included as an essential component of the present analysis.

We also note that, due to the complexity of the scalar potential arising in this model, a fully detailed dynamical analysis could not be carried out in complete generality. However, for specific and commonly used cosmological potentials, the dynamical framework developed here can be effectively applied to obtain a more detailed understanding of the model and to facilitate comparisons with standard scenarios.

In this respect, we have established two distinct formulations of the dynamical system for our model. 
 
The first one is the generalization of classical scalar field models, constructed through defining new variables. This system reduces to the corresponding classical models when $\alpha=1$. 
However, implementing this system into our fractional framework was difficult, since, unlike standard models, we did not introduce a simple hypothetical phenomenological potential by hand. Instead, we treated the potential as an unknown variable, resulting in a complicated function of $t$ that was nearly impossible to represent it as a function of the scalar field. However, as a generalization of standard models, it was included in the paper.

To establish the second formulation, the fractional equations were first rewritten as in the standard model, and then the corresponding dynamical system was easily constructed.
Since the resulting dynamical system is non-autonomous, we have introduced three different methods to analyze such systems and, in the end, adopted one of the most appropriate one in its most general form to study our fractional dynamical framework.

At this point, it is important to emphasize some fundamental aspects of the structure of our fractional model, which we now briefly discuss.

In standard cosmological frameworks, including canonical scalar field models, time diffeomorphism invariance is preserved, allowing for arbitrary reparametrizations of the time coordinate. However, in our present model, due to the presence of the time-dependent kernel in the action, the resulting field equations contain time--dependent terms that clearly break this symmetry. Nevertheless, one can still fix a specific gauge to simplify the dynamics.
Let us now be more precise. In our model, time plays a distinguished physical role, which may lead to significant consequences. Therefore, we should consider the following points.
The freedom to reparametrize time is lost, and the model becomes tied to a specific time foliation.
Important physical quantities such as the components of the EMT and scalar perturbations are most naturally interpreted within this chosen time frame.
Unlike standard covariant models, the breaking of this symmetry can give rise to additional dynamical behaviors that may not emerge otherwise, thereby offering new opportunities for describing various cosmological scenarios.
Concretely, the internal consistency of the theory must be carefully ensured, including the absence of ghost modes and the preservation of covariance at the perturbative level.
Importantly, the explicit breaking of the diffeomorphism invariance in extended frameworks like ours should not be regarded as a shortcoming. Rather, when properly controlled, this feature can be considered as introducing an extra degree of freedom that allows for a more flexible and geometrically motivated description of the universe without relying on ad hoc assumptions (such as adding ad hoc scalar potentials, as discussed in previous sections).

Concerning the Bianchi identities and the requirements for the effective EMT to be conserved, we briefly mention the following. It is generally accepted that, in any metric theory of gravity, local conservation of the EMT is derived from $\nabla_{\mu}G^{\mu\nu}\equiv0$. In our model, including the time-dependent kernel modifies the gravitational sector, see equation \eqref{SB-eq1}. Consequently, along with the continuity equation \eqref{eff-Con}, we can investigate whether the effective EMT remains divergence free under small departures from FLRW or breaks covariance; namely, the Noether second theorem (local symmetries).


It is worth noting that our model just incorporates a fractional time-dependent kernel into an extended SB model; not only does the scalar field have been assumed to be included in a kinetic term with the correct sign, but there are no higher-derivative terms or noncanonical structures. Therefore, the model is expected to be ghost--free both in the background and in the presence of linear perturbations.

It is important to make some comments on the fractional KG equation. As is well known, the KG equation is crucial in scalar field cosmologies for understanding the dynamical evolution of the universe. In our framework, the inclusion of fractional time--dependent terms, as indicated in equation \eqref{asli3}, causes this equation to gain a complex generalization in contrast to its standard version. 
There are two main approaches to interpreting the generalized Klein--Gordon equation in our framework:

\begin{enumerate}
    \item 
   
    \textbf{Effective friction interpretation:} The fractional time--dependent term can be absorbed into a generalized friction term, leading to an effective damping of the form
  $ \left[(n-1)H + \frac{1-\alpha}{t}\right]\dot{\phi}$.
    This equation demonstrates a frictional response that goes beyond the standard damping term.
    
    \item 
    
  \textbf{Effective potential approach:} As an alternative, one can redefine a time--varying effective potential $V_{\text{eff}}(\phi, t)$ so that the KG equation, along with potentially the other field equations, maintains a structure similar to its standard form. This method is particularly beneficial for directly analyzing the nature of the potential and comparing it with recognized potentials applied in models of inflation or dark energy.
\end{enumerate}

At this point, let us briefly highlight some of the restrictions imposed on our cosmological model.

\begin{itemize}

\item
In this study, while we have concentrated on the spatially flat FLRW model, it is clear that non-flat and anisotropic models, including various Bianchi types \cite{Rasouli:2011rv,rasouli2019extended,paliathanasis2022bianchi,chetia2023particle,Rasouli:2023idg,chokyi2024barrow,oliveira2024noncommutative,solanke2025lrs} and specifically the Kasner model \cite{Rasouli:2014sda,paliathanasis2018stability,paliathanasis2022kasner}, can also be analyzed within this framework.

\item 
In future research, a more comprehensive analysis of effective potential might reveal new characteristics of the model. Specifically, contrasting the resulting $V_{\text{eff}}(\phi, t)$ with common potentials such as $m^2\phi^2$, $\lambda\phi^4$, or plateau--like potentials such as Starobinsky's model could lead to the development of creative scalar field scenarios influenced by the foundational fractional structure.

\end{itemize}
Regarding the proposal of the extended scenarios mentioned, there are potential pathways to achieve such intriguing solutions. However, investigating these generalizations and evaluating their physical feasibility will be integral to our future research plans within the framework of current or other fractional models.

\begin{acknowledgments}
We are grateful to the anonymous referees for their insightful comments and constructive suggestions, which significantly improved the clarity and quality of this work. SMMR and PM acknowledge the FCT grant \textbf{UID/212/2025} Centro de Matem\'{a}tica 
e Aplica\c{c}\~{o}es da Universidade da Beira Interior plus
the COST Actions CA23130 (Bridging high and low energies in search of
quantum gravity (BridgeQG)) and CA23115 (Relativistic Quantum Information (RQI)).
\end{acknowledgments}

\bibliographystyle{unsrt} 
\bibliography{SBFrac}

\end{document}